\documentstyle[12pt,psfig]{article}

\newcommand{\beq}{\begin{equation}}
\newcommand{\eeq}{\end{equation}}
\newcommand{\bea}{\begin{eqnarray}}
\newcommand{\eea}{\end{eqnarray}}

\newcommand{\gsim}{\lower.7ex\hbox{$
\;\stackrel{\textstyle>}{\sim}\;$}}
\newcommand{\lsim}{\lower.7ex\hbox{$
\;\stackrel{\textstyle<}{\sim}\;$}}

\def\lsim{\mathrel{\rlap{\lower3pt\hbox{\hskip0pt$\sim$}}
    \raise1pt\hbox{$<$}}}         %less than or approx. symbol
\def\gsim{\mathrel{\rlap{\lower4pt\hbox{\hskip1pt$\sim$}}
    \raise1pt\hbox{$>$}}}         %greater than or approx. symbol

\newcommand{\bibit}[1]{\bibitem{#1}}

\newcommand{\aver}[1]{\langle #1\rangle}

\newcommand{\La}{\overline{\Lambda}}
\newcommand{\Lam}{\Lambda_{\rm QCD}}
\newcommand{\mhad}{\mu_{\rm hadr}}

\newcommand{\as}{\alpha_s}
\newcommand{\GeV}{\,\mbox{GeV}}
\newcommand{\MeV}{\,\mbox{MeV}}
\newcommand{\matel}[3]{\langle #1|#2|#3\rangle}

\newcommand{\gsl}{\Gamma_{\rm sl}(b\!\to\!c)}
\newcommand{\asMS}{\alpha_s^{\overline{\rm MS}}}

\newcommand{\msp}[1]{\mbox{\hspace*{#1mm}~}}

\begin{document}
% ------------- Titlepage -------------- 
\thispagestyle{empty}
\begin{flushright}
Bicocca-FT-03-5\\
TTP03-07 \\
UND-HEP-02-BIG\hspace*{.08em}08\\
hep-ph/0302262\\
%%% {\tiny newvcbhep.tex \today }\\
\end{flushright}
\bigskip
\boldmath
\begin{center}
{\Large{\bf Imprecated, yet Impeccable:\vspace*{5mm}\\
On the Theoretical Evaluation of 
$\,\Gamma (B \!\to\! X_c \,\ell \nu)$}}\vspace*{9mm} 

{\tt Submitted to the Proceedings of the CKM Workshop}
\vspace*{7mm} 
\end{center}

\unboldmath
\smallskip
\begin{center}
{\large{D.~Benson$^{\,a,b}$, I.I.~Bigi$^{\,a,b}$, 
Th.~Mannel$^{\,b}$ and N.~Uraltsev$^{\,c\,*}$
}}  \\
\vspace{4mm}
$^a$ {\sl Department of Physics, University of Notre Dame du Lac}
\vspace*{-.8mm}\\
{\sl Notre Dame, IN 46556, USA}$^{\,1}$\vspace*{.5mm}\\
$^b$ {\sl Institut f\"{u}r Theoretische Teilchenphysik, 
Universit\"{a}t Karlsruhe}\vspace*{-.8mm}\\
{\sl D--76128 Karlsruhe, Germany}\vspace*{.5mm}\\ 
$^c$ {\sl INFN, Sezione di Milano, Milan, Italy} 
\vspace*{12mm}

{\bf Abstract}\vspace*{-.9mm}\\
\end{center}
\noindent
We present a detailed evaluation of the total semileptonic 
$B$ meson width in terms of $|V_{cb}|$ and heavy quark parameters 
(quark masses and the expectation values of local heavy quark operators).
Special attention is given to perturbative corrections which can 
precisely be calculated in a scheme with a 
hard Wilsonian cutoff at a scale around $1\GeV$ appropriate for 
the OPE, and to the 
potential impact of higher-order power corrections. We point out that
the  latter require control
over possible contributions from four-quark operators containing charm
quark fields. Analytical expressions are given which allow evaluating the
width with various choices of parameters; ready-to-use  expressions
showing the dependence on the heavy quark parameters are presented as
well. We illustrate these results by commenting on how  these parameters
can be extracted  and what accuracy is likely to be achievable in the near
future. 

\setcounter{page}{0}

\vfill

~\hspace*{-12.5mm}\hrulefill \hspace*{-1.2mm} \\
\footnotesize{%% \noindent 
\hspace*{-5mm}$^1$Permanent address\\
\hspace*{-5mm}$^*$On leave of absence from Department of 
Physics, University of Notre Dame, Notre Dame, 
IN 46556, USA \hspace*{-7mm}~\\
\hspace*{-5mm}\hspace*{.25em} and St.\,Petersburg Nuclear Physics 
Institute, Gatchina, St.\,Petersburg  188300, Russia}

\newpage
\tableofcontents
\vspace{1cm}

% ------------------- End Titlepage -------------------
%\newpage
%
%
%
%
Extracting the value of $|V_{cb}|$ from the observed semileptonic width 
of $B$ mesons constitutes a method of impeccable theoretical 
pedigree \cite{rev}, although it often has been imprecated in the past. 
The theoretical part of the analysis proceeds in two steps: 
\begin{enumerate}
\item 
Through a heavy quark expansion (HQE) one expresses 
$\Gamma (B \to l \nu X_c)$ as $|V_{cb}|^2$ times a function of 
the heavy quark (HQ) parameters  
like quark masses and hadronic expectation values, and the 
perturbative contributions.
\item 
One determines the values of these HQ parameters from observables other 
than the total semileptonic width. 
\end{enumerate}
The first step has now reached a mature stage 
after having been subjected to considerable scrutiny for 
several years. The 
primary motivation for this note is to present an updated 
description of this first task. We will give analytical 
expressions as much as reasonably possible, state explicitly 
the proper field theoretical definitions of the HQ 
parameters and 
address the potential sources for theoretical uncertainties that 
arise in such an expansion. We aim at giving an 
`open source code' which enables 
the dedicated reader to check the validity of the
theoretical results and insert personally preferred values for the 
HQ parameters and incorporate additional constraints.

To extract a value $|V_{cb}|$ from the observed width one has, of course 
to complete step two as well. There  
exists considerable information and constraints on these HQ 
parameters. We will sketch how they can be determined. However, 
more work is still needed. We will present numbers in this
context; yet they are to be understood as illustrative rather than
final.

There is already highly nontrivial evidence  that the systematics are
indeed under control. We anticipate that in the  end one will be able to
reduce the overall theoretical  uncertainty to the one percent level. This
benchmark will guide us in calculating  $\Gamma_{{\rm sl}}(B)$.

The remainder of this note will be organized as follows: 
in Sect.~\ref{EXSUM}
we provide a summary of the expressions that one can use to obtain the value
of $|V_{cb}|$ from the total semileptonic $B$ width. 
In Sect.~\ref{GAMMA} we present the theoretical expressions 
used to derive the total semileptonic $B$ width and discuss all
ingredients in the corresponding subsections. We include the effects of
four-quark operators with charm field previously ignored, and analyze the
potential effect of higher order power corrections. 
Then we discuss in more detail the definitions and
basic  properties of the heavy quark operators, and address the 
perturbative corrections in Sect.~\ref{PERTCONT}. 
Section~\ref{UNCERT} summarizes the theoretical uncertainties in evaluating
$\Gamma_{\rm sl}$.  In Sect.~\ref{HQP} we illustrate how one can
independently determine the values of the required heavy quark parameters
from experiment. Technical details are relegated to Appendices, where
we give explicit expressions for the normalization point dependence 
of the heavy quark parameters, provide formulae for computing
perturbative corrections implementing Wilsonian OPE, and briefly 
describe the BLM summation to arbitrary order within this framework.

%%%%%%%%%%%%%%
\section{Executive Summary} 
\label{EXSUM}
%%%%%%%%%%%%%%%

We base our numerical 
analysis on the following expression for the semileptonic $B$
width \cite{prl} through order
$1/m_Q^3$:
\small
\bea
\nonumber
\Gamma_{\rm sl}(b\!\to\!c)\msp{-3} &=& \msp{-3}\frac{G_{F\,}^2
m_b^5(\mu)}{192\,\pi^3}\; \raisebox{-.5mm}{\mbox{{\large$|V_{cb}|^2$}}} 
\,\left(1\!+\!A_{\rm ew}\right)\,A^{\rm pert}(r,\mu)\!\left[z_0(r)\!\left(
1\!-\!\frac{\mu_\pi^2(\mu)\!-\!\mu_G^2(\mu) \!+\! 
\frac{\rho_D^3(\mu)\!+\!\rho_{LS}^3(\mu)}{m_b(\mu)}}
{2m_b^2(\mu)}\right)\right.\\
&-& \msp{-3} \left. 2(1\!-\!r)^4 
\frac{\mu_G^2(\mu)\!-\!\frac{\rho_D^3(\mu)\!+
\!\rho_{LS}^3(\mu)}{m_b(\mu)}}{m_b^2(\mu)}+
d(r)\,\frac{ \rho_D^3(\mu)}{m_b^3(\mu)}+ \!...
\right]\!,
\qquad \qquad 
\label{SLwid}
\eea
\normalsize
where $z_0(r)$ is the tree-level phase space factor and 
$r\!=\!m_c^2(\mu)/m_b^2(\mu)$:
\beq
z_0(r)=1-8r+8r^3-r^4-12r^2\ln{r}\;,
\label{8}
\eeq
while the expression for $d(r)$ follows from Ref.~\cite{grekap}:
\beq
d(r) =  8\ln{r} + \frac{34}{3}-\frac{32}{3}r -8 r^2
+\frac{32}{3}r^3 -\frac{10}{3} r^4 \;\simeq\; -18.3 \,z_0 \;\mbox{ at }
\sqrt{r}\!=\!0.25\;.
\qquad 
\label{9}
\eeq
The electroweak correction, $A_{\rm ew}$ corresponding to the ultraviolet
renormalization of the Fermi interaction is well-known \cite{ewlog}: 
\beq
1+A_{\rm ew}\simeq 
\left(1+\frac{\alpha}{\pi} \,\ln{\frac{M_Z}{m_b}}\right)^2\simeq 1.014.
\label{Aew}
\eeq 
An auxiliary scale $\mu$ 
is introduced to demark the border between long- and short-distance 
dynamics in the OPE. Unless stated otherwise, we adopt $\mu \simeq 1 $ GeV.

The quantity $A^{\rm pert}$ accounts for the perturbative
contributions; it has been calculated  
to all orders in BLM corrections and to second order in non-BLM 
corrections as discussed in detail in Sect.~\ref{GAMMA}.  
For $\sqrt{r}\!=\!0.25$ and $\mu=1 \GeV$ we find that 
$A^{\rm pert}\simeq 0.908$.

The quantities $\mu_{\pi}^2$, $\mu_G^2$, $\rho_D^3$ and $\rho_{LS}^3$  
denote the expectation values of the kinetic, chromomagnetic, 
Darwin and spin-orbit operators, respectively. 
Only $\rho_D^3$  
has a noticeable impact at ${\cal O}(1/m_b^3)$. In that order in $1/m_b$
there arise also contributions corresponding to four-quark operators 
of the generic form $(\bar b \Gamma c)(\bar c\Gamma b)$. Their 
nonperturbative 
$B$ expectation value, while small, will not vanish. 
For now we include such a contribution 
in our estimate of the theoretical uncertainty, given below.

Using the numerical values $m_b(1\GeV)\!=\!4.6\GeV$,
$\mu_\pi^2(1\GeV)\!=\!0.4\GeV^2$, $\mu_G^2(1\GeV)\!=\!0.35\GeV^2$,  
$\rho_D^3(1\GeV)\!=\!0.2\GeV^3\,$ and 
$\,\rho_{LS}^3(1\GeV)\!=\!-0.15\GeV^3\:$, for $\:\sqrt{r}\!=\!0.25\,$ and
$\,\alpha_s(m_b)\!=\!0.22\,$ we find the following 
expression for $|V_{cb}|$:\footnote{The number slightly differs from
Ref.~\cite{blmvcb} since we adopt the Wilsonian renormalization
convention for Darwin term as well; our $\rho_D^3(\mu)$ should be
distinguished from $\tilde\rho_D^3$ of Refs.~\cite{blmvcb,DELPHI}.} 
\beq
\frac{|V_{cb}|}{0.0417}\simeq \; \mbox{{\large$
\left(\frac{{\rm Br}_{\rm sl}(B)}{0.105}\right)^{\!\frac{1}{2}} 
\left(\frac{1.55{\rm ps}}{\tau_B}\right)^{\frac{1}{2}}$}}
\left(1\!-\!\mbox{{\small$4.8\,[{\rm Br}(B\!\to\!X_u\,\ell\nu)-0.0018]$}}
\right)\, (1+\delta _{\rm th}) 
,
\label{vcb1}
\eeq
with the following dependence of $|V_{cb}|$ on the various heavy 
quark parameters:
{\small
\begin{eqnarray}
\frac{|V_{cb}|}{0.0417} \msp{-3}&=&\msp{-3} 
(1+\delta _{\rm th})\:
[1+0.30\,(\alpha_s(m_b)\!-\!0.22)]\, \times \nonumber 
\\ 
& & \msp{22}
\left[1 -0.66 \left ( m_b(1\GeV) \!-\!4.6\GeV \right )
+0.39 \left ( m_c(1\GeV) \!-\!1.15\GeV \right )\qquad \right. \nonumber \\
& & \msp{26}+0.013 \left ( \mu_{\pi}^2 \!-\!0.4\GeV^2 \right )
+0.09 \left ( \rho_D^3 \!-\!0.2\GeV^3 \right ) \nonumber \\
& & \msp{26}\left. +\,0.05 \left ( \mu_G^2\!-\!0.35\GeV^2 \right ) 
-0.01 \left ( \rho_{LS}^3 \!+\!0.15\GeV^3 \right ) \right ] .
\label{vcb2}
\end{eqnarray}
}

Beyond the uncertainties in the numerical values of these parameters
there is the potential error in $|V_{cb}|$ due to the limited accuracy
of the
theoretical expression for the semileptonic width we relied upon; this is 
denoted by $\delta_{\rm th}$, for which we estimate 
\beq
\delta_{\rm th}=\pm 0.005_{\rm pert}\pm 0.012_{\rm hWc}
\pm 0.004_{\rm hpc} \pm 0.007_{\rm IC}\;.
\label{sumunc}
\eeq
The terms here represent the remaining uncertainty in the Wilson 
coefficient of the leading operator 
$\bar bb$ (perturbative correction), as yet
uncalculated perturbative corrections to the Wilson coefficients of 
chromomagnetic and Darwin operators,  
higher-order power corrections including violation of local duality in
$\gsl$,
and possible nonperturbative effects in the operators with
charm fields, respectively.

While the numbers of Eq.~(\ref{sumunc}) are small, we consider them 
rather on the conservative side. Yet they should be viewed as 
what they are, namely estimates based on best efforts rather 
than mathematical theorems.
 
%%%%%%%%%%%%%%
\section{\boldmath $\Gamma (B\to X_c\,\ell \nu)$; \,\,Master Formulae} 
\label{GAMMA}
%%%%%%%%%%%%%%%

The operator product expansion (OPE) yields 
inclusive heavy quark decay rates as an asymptotic series in inverse 
powers of the heavy quark mass. 
To be more precise, there are different mass scales in the problem;
the most important one is the energy release given by 
$E_r= m_b \!-\! m_c$ for $b\to c$ 
\cite{rev}. Hence, the width in the OPE  
is an expansion in inverse powers of $E_r$.\footnote{While the 
OPE in Eqs.~(\ref{SLwid}), (\ref{10}) at first sight look like 
expansion in powers of $1/m_b$, this is not really the case. In the 
SV limit $m_b \!-\! m_c \ll m_b$ the expansion 
parameter is obviously $1/(m_b - m_c)$; at $m_b \!\to\! m_c$ 
the expansion becomes meaningless no matter how large $m_b$ is. 
Constructing the OPE directly for the widths shows that the expansion 
parameter in general is 
$1/(m_b\!-\! m_c)$ \cite{ioffe}, which comes about in
Eqs.(\ref{SLwid}), (\ref{10}) due to the interplay of $1/m_b$ and 
$r\!=\!m_c^2/m_b^2$. 
On the other hand expectation values like $\matel{B}{\bar bb}{B}$ 
are expanded in powers of $1/m_b$.} 
It is useful to note that for actual quark masses 
the dependence on $m_b$ and $m_c$ 
is well approximated by $\Gamma_{{\rm sl}}(B) 
\propto m_b^2\,(m_b \!-\! m_c)^3$.

Through order $1/m_Q^3$ one has the general expression 
\small
\bea
\nonumber
\Gamma_{\rm sl}(b\!\to\!c)\msp{-3.7} &=& \msp{-3.7}\frac{G_{F\,}^2
m_b^5(\mu)}{192\,\pi^3}\; \raisebox{-.5mm}{\mbox{{\large$|V_{cb}|^{2\!}$}}} 
\,\left(1\!+\!A_{\rm ew}\right)\!\left[ z_0(r)\,
[1\!+\!A_3^{\rm pert\!}(r;\mu)]\!\left(\!
1\!-\!\frac{\mu_\pi^2(\mu)\!-\!\mu_G^2(\mu) \!+\! 
\frac{\rho_D^3(\mu)\!+\!\rho_{LS}^3(\mu)}{m_b(\mu)}}
{2m_b^2(\mu)}\right)\right.\\
&-& \msp{-3} \left. 
(1+A_5^{\rm pert}(r;\mu))\, 2(1\!-\!r)^4 
\frac{\mu_G^2(\mu)\!-\!\frac{\rho_D^3(\mu)\!+
\!\rho_{LS}^3(\mu)}{m_b(\mu)}}{m_b^2(\mu)}+
(1+A_D^{\rm pert}) d(r)\,\frac{ \rho_D^3(\mu)}{m_b^3(\mu)}
\right.\nonumber\\
&+& \msp{-3} \left.
32\pi^2\,(1+A_{6c}^{\rm pert}(r))(1\!-\!\sqrt{r})^2
\frac{H_c}{m_b^3(\mu)} + 
32\pi^2\,\tilde A_{6c}^{\rm pert}(r) 
(1\!-\!\sqrt{r})^2\frac{\tilde H_c}{m_b^3(\mu)}
\right.\nonumber\\
&+& \msp{-3} \left.
32\pi^2 A_{6q}^{\rm pert}(r)\frac{F_q}{m_b^3(\mu)} +
{\cal O}\left(\frac{1}{m_b^4}\right)
\right]\!,
\qquad \qquad 
\label{10}
\eea
\normalsize
The phase space factor $z_0(r)$ has been defined in Eq.~(\ref{8}) and
the tree coefficient for the Darwin term $d(r)$ in Eq.~(\ref{9}).
A number of clarifications are to be made about the above master expression.

Following Wilson's prescription for the OPE an auxiliary scale $\mu$ 
has been introduced to separate the effects from long- and short-distance
dynamics.  Since observables do not depend on it, quark masses,
radiative  corrections and hadronic expectation values 
$\matel{B}{O_i}{B}$ combine to yield a compensating 
$\mu$ dependence as indicated in Eq.~(\ref{10}). 
The natural and most efficient 
choice is $\mu \simeq 1\GeV$, which is particularly relevant for the
quark masses \cite{five}.

It should be kept in mind that the $B$ state over which the 
expectation values are taken is the real $B$ meson, not the meson  
in the heavy quark limit $m_b \to \infty$. In this convention the
total width to order $1/m_b^3$ is corrected by a single Darwin
expectation value (modulo the effect of four-quark operators to be
addressed below). No non-local correlators enter, and the
spin-orbital expectation value $\rho_{LS}^3$ enters only as a
$1/m_b$ piece of the total Lorentz-scalar chromomagnetic expectation
value \cite{optical}.\footnote{The inclusive moments of various decay
distributions do exhibit explicit dependence on $\rho_{LS}^3$;
it is however too weak to matter in practice.} Likewise, the kinetic
expectation value enters solely through the $1/m_b$ expansion of the
leading operator $\bar{b}b$ \cite{buv,prl,optical}.

All Wilson coefficients are given by short-distance dynamics and in
practice are evaluated in perturbation theory. In the convention adopted
in  Eq.~(\ref{10}) the tree-level coefficients $z_0$, $d$ etc.\
are modified by perturbative corrections $A_{...}^{\rm pert}$
starting with at least one power of $\alpha_s$.

The leading nonperturbative corrections arise in order $1/m_Q^2$ and 
are controlled by the expectation values 
$\mu_{\pi}^2(\mu)$  and $\mu_{G}^2 (\mu)$ of the  kinetic and chromomagnetic 
dimension-five operators ($D_{\nu}$ denotes the covariant derivative), 
respectively 
\beq
\mu_{\pi}^2(\mu )  \equiv \frac{1}{2M_B}
\langle B|\overline{b}(i\vec D)^2 b|B \rangle_\mu\;,\qquad
%\textstyle
\mu_{G}^2 (\mu )  \equiv \frac{1}{2M_B}
\langle B|\overline{b}\frac{i}{2}
  \sigma_{jk} G^{jk}b|B \rangle_\mu\,.
\label{16} 
\eeq
(Since we have explicitly 
introduced the effect of the $LS$ operator, $\mu_G^2$ above should be
understood as the average of only the chromomagnetic field,
$\aver{\vec\sigma \vec B}$, although it includes
$1/m_b$ finite-mass effects.) 
The Darwin and spin-orbital $LS$ terms $\rho_{D}^3(\mu )$ and 
$\rho_{LS}^3(\mu )$ emerge from dimension-six operators: 
\beq
\rho_{D}^3(\mu )  \equiv \frac{1}{2M_B}
\langle
B|\overline{b}(-\frac{1}{2}
  \vec{D} \!\cdot\! \vec{E})b|B \rangle _\mu \;,\qquad
%\textstyle
\rho_{LS}^3(\mu )  \equiv \frac{1}{2M_B}
\langle
B|\overline{b}(\vec{\sigma}\!\cdot\!
  \vec{E}\! \times \!i\vec{D})b|B \rangle_\mu \;.
\label{18} 
\eeq
The Wilson coefficients for the operators describing nonperturbative 
effects are completely known only at tree
level. Since they represent power-suppressed effects showing up at 
the one
percent level, this is a reasonable approximation. However, it is
desirable to improve it in the future computing ${\cal O}(\alpha_s)$
corrections to them.

The last term in Eq.~(\ref{10}) proportional to $F_q$ denotes the effect of
generic $SU(3)$-singlet four-quark operators, other than Darwin
operator, of the form
$\bar{b}\Gamma b\:\bar{q}\Gamma q$ with the sum over $q=u,d,s$, and
$\Gamma$ including both color and Lorentz matrices (to the leading
order in $1/m_b$ only $\gamma_0\times \gamma^0$ or
$\gamma_i\gamma_5\times\gamma^i\gamma_5$ structures survive, but one does
not need to rely on this). Their Wilson coefficients are order
${\cal O}(\alpha_s)$, and we neglect these contributions.

On the other hand, we have explicitly included in the master equation a
possible effect of the (tree level) expectation values of the
four-quark operators with the {\tt charm} field. It is routinely
skipped assuming that it vanishes due to the sizeable charm mass. While
it represents a reasonable approximation, it can be valid only up to a
certain accuracy. Since the Wilson coefficient of one linear combination of
four (in the heavy-$b$ limit) such operators emerges at tree
level and is strongly enhanced by the effectively two-body phase
space, this effect might not be totally
negligible. To complement its purely theoretical estimates or
upper bounds, we find it advantageous to explicitly introduce this
effect and study its possible manifestations in experiment:
\beq
H_c=\frac{1}{2M_B}
\matel{B}{\bar{b}\gamma_\nu(1\!-\!\gamma_5)c\, 
\bar{c}\gamma^\rho(1\!-\!\gamma_5)b}{B}_\mu \: 
\left(-\delta^\nu_\rho\!+\!v^\nu v_\rho\right)\,, \qquad 
v_\nu = \frac{P^B_\nu}{M_B}\;.
\label{24}
\eeq
 This way to refer to such nonperturbative effects literally
applies when normalization point $\mu$ is taken above $m_c$.
Having the normalization scale $\mu$ below $m_c$ amounts to the 
$c$ quark field being integrated out; then there are no dynamical 
$c$ quark fields. 
Nevertheless, the physical effects 
lumped into $H_c$ still 
do not have to vanish exactly; they are rather partially reshuffled
into higher-dimension b-quark operators with only light fields.

The second charm-related $D\!=\!3$ parameter $\tilde H_c$ in
Eq.~(\ref{10}) represents the cumulative effect of the remaining
combinations of the four-fermion operators with charm quarks. It is
expected to play an insignificant role, but will be briefly addressed as
well.

It turns out that the consistent $1/m_b$ expansion to higher orders 
{\tt mandates} introducing such four-quark charmed operators. They
can be excluded only at the price of spoiling the expansion: instead
of $1/m_b$ or $1/E_r$ some of the higher-dimensional operators would get
powers of $m_c$ in the denominator to compensate for the increasing
dimension. To state it differently: the Wilson coefficients 
of the higher-dimensional operator can contain an 
enhancement by positive powers of
$1/r$. The aggregate effect of such terms can be viewed as 
the definition of $H_c$ at low normalization scale.
Moreover, a significant fraction of the numerically enhanced Darwin 
coefficient function $d(r)$ represents actually the mid-virtuality
(semi-hard, or hybrid) contribution to the leading charm operator.
A detailed discussion of these operators and their relation to 
the notion of 
``Intrinsic Charm'' will be given in a separate publication
\cite{ic}. We shall briefly recapitulate the main points of that
analysis below in Sect.~\ref{ic}.

The Wilsonian separation of ``soft'' and ``hard'' effects can be done
in different ways. In particular, this affects the precise values of
the heavy quark masses at a given $\mu$. We consistently use the
scheme based on the Small Velocity, or Shifman-Voloshin (SV) sum rules
which introduce the normalization scale $\mu$ via the cutoff over the
excitation energy of the hadronic states \cite{five,blmope}. 
Other
possibilities to define heavy quark masses discussed in the literature 
are considered in the review
\cite{hoang}.\footnote{Their assessment in Ref.~\cite{hoang} is not
always justified in our opinion.} The
advantages of this choice are transparent. First, they allow to define 
the heavy quark masses as well as the expectation values of the 
higher-dimensional operators on
a parallel footing. Secondly, being expressed via (in principle)
measurable quantities, they have a definite value in  
QCD at any given normalization point. The question of their precise
extraction from experiment then becomes meaningful (see, e.g.\
Ref.~\cite{chrom}),  without invoking doubtful procedures. Thirdly, such a
renormalization scheme respects exact inequalities established 
in the heavy quark limit (for a
recent review, see Ref.~\cite{ioffe}), which turn out very
constraining in practice and allow to go far beyond naive dimensional
estimates even for higher-dimensional operators. Finally, it has been
demonstrated in numerous applications that using such a scheme typically
yields a much improved convergence of the perturbative expansion, see,
e.g.\ a recent experimental implementation in Ref.~\cite{DELPHI}.

Below we discuss the salient features in more detail.

%%%%%%%%%%%%%%%%
\section{Heavy Quark Parameters}
\label{HQPARAM}
%%%%%%%%%%%%

%%%%%%%%%%%%%%
\subsection{Heavy quark masses}
\label{QUARKMASS}
%%%%%%%%%%%%%%%

With the width depending on the fifth power of the heavy quark mass any
uncertainty associated  with it -- conceptual or numerical -- 
has a grave
impact on the  accuracy with which $|V_{cb}|$ can be extracted.
Therefore great care has to be applied to properly defining it and
consistently calculating the radiative corrections.

The {\em pole mass} is defined 
in analogy to the electron case in QED as the position of the pole 
in the quark Green function. This pole mass is gauge invariant 
and perturbatively infrared finite. 
Employing it is often 
convenient for purely  perturbative calculations. 
Yet in full QCD due to confinement there is no such pole -- 
the pole mass is thus not `infrared stable' due to nonperturbative 
dynamics. Moreover, it has an intrinsic
and thus irreducible  theoretical uncertainty $\sim\!\Lam$
already in perturbation theory; this
makes it inappropriate when one wants to include nonperturbative
contributions, which are  power suppressed \cite{pole}: 
\beq 
\Gamma_{\rm sl}(B) \propto m_b^5 \simeq (m_b^{\rm pole})^5 
(1 + 5{\cal O}(\Lam)/m_b) \; ; 
\label{40}
\eeq
i.e., the intrinsic uncertainty is at least parametrically larger than 
the leading nonperturbative contributions that arise in order 
$1/m_b^2$. The meaning of this observation can be easily understood. 
A long-distance pole mass in perturbation theory is a counterpart of 
the heavy flavor {\em hadron}\, mass of full QCD, since it includes
effects of gluons with arbitrarily small momenta. In this respect the
analogue of the quark mass in QCD where nonperturbative
effects exit, is a short-distance mass $m_Q(\mu)$ of perturbation
theory, since such a mass excludes soft gluons interacting strongly.
The central OPE result that there are no 
contributions $\sim {\cal O}(1/m_b)$ contributions to fully inclusive 
widths \cite{buv,bs} -- in other words, the hadron ($B$ meson) mass
is irrelevant for the width which depends only on $m_b$, but not on 
$M_B\!-\!m_b$ -- 
translates then into the statement 
that short-distance masses have to be used when aiming at the 
power-like accuracy.

To a limited extent, the pole mass can still be used to approximate
the width: the order-to-order shift in the apparent position of the 
pole quark mass in perturbation theory is offset by a significant
change in the overall perturbative correction factor. Yet such a 
procedure may not allow to properly account already for the actual
leading nonperturbative corrections. This is illustrated by the
following simple observation. 
Starting with the short-distance mass, one
has for the leading (parton) contribution  
\beq 
\Gamma_{\rm sl}(B) \propto A^{\rm pert}(\mu) \,m_b^5(\mu) 
\label{42}
\eeq
with neither factor having uncontrollable infrared pieces. The same
perturbative expression for the pole masses then must be of the form
\beq
\Gamma_{\rm sl}(B) \propto A^{\rm pert}(\mu) 
(1 \!-\!5\mbox{$\frac{\tilde\Lambda}{m_b}$}) \,(m_b^{\rm pole})^5 
\simeq A^{\rm pert}(\mu) \,m_b^5(\mu) 
\left[1-15\left(\mbox{$\frac{\tilde\Lambda}{m_b}$}\right)^2\right], 
\label{44}
\eeq
where $\tilde\Lambda/m_b$ expresses the uncontrollable infrared
contributions in the ratio of the pole to short-distance mass. 
The contribution $\sim \!{\cal O}(1/m_b)$ has indeed disappeared, yet 
the expression is still  deficient: while there are contributions 
$\sim \!{\cal O}(1/m_b^2)$, they have to be of the form 
$1-\frac{1}{2}\frac{\mu_{\pi}^2}{m_b^2}$ regardless of the particular 
interactions.\footnote{The perturbative contributions discussed here
cannot give rise to spin-dependent effects contained in $\mu_G^2$.} 
Hence the last term in Eq.~(\ref{44}) coming with the
coefficient $15$ has no counterpart in actual QCD and is rather an 
artefact of using the pole masses.

An example of a short-distance $b$ mass is the 
$\overline{{\rm MS}}$ mass $\bar m_b(m_b)$. The
normalization scale $m_b$, however is unnaturally high for the problem; 
therefore it introduces significant higher-order corrections not related
to running of $\alpha_s$ per se, which are under poor control and thus
limit the accuracy \cite{five}.

A better choice when treating $B$ decays is taking the Wilsonian 
factorization scale $\mu$ around $1\GeV$. An example of such a mass
is the one defined through the SV sum rules, the prescription valid to
arbitrary perturbative order. It has the meaning of the mass entering
the kinetic energy of the heavy quark and is often  called {\it kinetic
mass}\, $m_b^{\rm kin}(\mu)$. At every given $\mu$ it represents an
observable physical quantity and, with an appropriate choice of $\mu$ 
its extraction from experiment provides good stability with respect to
radiative corrections. Its definition is discussed in
Sect.~\ref{apmQ}. Here we mention that its value has been extracted from
the threshold domain of the $e^+e^-$ annihilation into $\bar{b}b\,$, 
\beq
m_b^{\rm kin}(1\GeV)\simeq (4.57\pm 0.06)\GeV
\label{48}
\eeq
  From now on we omit the superscript ``kin'' using this mass as the
default definition.

For the charm quark, the separation scale taken around $1\GeV$ is 
similar in magnitude to $m_c$ itself. The kinetic mass $m_c(\mu)$
can still be used, with the realization, however, that this definition
starts loosing direct physical meaning for increasing $\mu$.

Historically heavy use has been made of relating the quark mass 
difference $m_b\!-\!m_c$ to the spin-averaged meson masses 
to reduce the number of parameters and to suppress ambiguities related to
heavy mass definitions: 
\beq 
m_b - m_c = \bar M_B - \bar M_D  
+ \frac{\mu _{\pi}^2}{2} 
\left( \frac{1}{m_c} - \frac{1}{m_b} \right) 
+ \frac{\rho _D^3 - \bar \rho ^3}{4}
\left( \frac{1}{m_c^2} - \frac{1}{m_b^2} \right) 
+ {\cal O}(1/m_Q^3) 
\label{MBMC}
\eeq
where $\bar M_{B[D]}  \equiv M_{B[D]}/4 + 3M_{B^*[D^*]}/4$ denotes the
spin averaged meson masses  and $\bar \rho ^3 \!=\! \rho _{\pi\pi}^3 +
\rho_s^3$ the sum of two  positive (in the advocated scheme) nonlocal
correlators (they are defined in Ref.~\cite{optical}). 
Eq.~(\ref{MBMC}) has become a widely used relation: for once one 
realizes that $\mu_{\pi}^2$ cannot differ from $0.45\GeV^2$ by 
significantly more than $0.1\GeV^2$, it leads to a reduced uncertainty 
in $m_b\!-\!m_c$. Having
the same scheme and normalization point for both $m_b$ and $m_c$ has
an advantage of making the heavy quark expansion relation 
between $m_b\!-\!m_c$ and $M_B\!-\!M_D$ more direct. 
If such constraints are
not imposed, it is justified to use $\bar m_c(m_c)$ as well.
We argue in Sect.~6 that it is safer not to invoke these relations. 

%%%%%%%%%%%%%%%%%
\subsection{Operators of dimension \boldmath $5$ and $6$}
%%%%%%%%%%

The precise definition of higher-dimensional operators affects
perturbative corrections to a lesser extent than the choice of 
the heavy quark
mass scheme. Yet it is required to make sense of assigning them a definite
numerical value; once again the values have to be 
normalization-scale dependent. In particular this refers to the
spin-singlet expectation values $\mu_\pi^2$ and $\rho_D^3$. They can be
defined in the same physical way as the heavy quark masses; the formal
definition is given, e.g.\ in \cite{chrom}. Once given, it allows
to sensibly extract their values; for instance, Ref.~\cite{chrom}
concluded from the hyperfine mass splitting that 
\beq
\mu_G^2(1\GeV)=(0.35^{+.03}_{-.02})\GeV^2\;.
\label{60}
\eeq
Kinetic expectation value $\mu_\pi^2$ a priori is less certain,
however in this regularization scheme the inequality 
\beq
\mu_\pi^2(\mu) \ge \mu_G^2(\mu)
\label{62}
\eeq
holds for {\tt any} normalization scale. With a number of experimental
data strongly favoring $\mu_\pi^2(1\GeV)\lsim 0.5\GeV^2$ a limited
range is left allowed, and the estimate 
\beq
\mu_\pi^2(1\GeV))=(0.45\pm 0.1)\GeV^2
\label{64}
\eeq
seems rather conservative. It is also supported by purely theoretical
estimates including QCD sum rules \cite{bloklub}.\footnote{According 
to the FNAL group,
lattice studies suggest so far a somewhat larger value \cite{kronsim}.}

A less 
precise estimate can
be obtained for the Darwin expectation value, $\rho_D^3(1\GeV)\simeq
(0.2\pm 0.1)\GeV^3$, yet it must be positive. The spin-orbit
average $\rho_{LS}^3$ is expected to be negative. It obeys
the constraints
\beq
-\rho_{LS}^3 \le \rho_{D}^3, \qquad |\rho_{LS}^3| \le 2 \rho_{D}^3\;.
\label{68}
\eeq
Since the effect of this operator is typically strongly suppressed
compared to the Darwin term, already these general bounds 
suffice for practical purposes.

A clarification is in order here. Strictly speaking, all the above
bounds hold in the heavy quark limit, i.e.\ when $1/m_b$ corrections
in the expectation values are absent. It has been argued \cite{chrom}
that their shift from the $m_b\!\to\!\infty$\, limit is negligible in
$B$ meson. Not
only it is governed by the parameter $1/2m_b$, it should be
additionally suppressed by virtue of the small excess of $\mu_\pi^2$
over $\mu_G^2$ manifesting proximity to the so-called BPS limit for
the heavy meson ground state. Then the deviation of these hadronic
parameters  lies well below available precision, and can be
neglected in practice.

%%%%%%%%%%%%%%%%%
\subsection{Operators with charm quarks} 
\label{ic}
%%%%%%%%%%%

As mentioned in the Executive Summary and the beginning of 
Sect.~\ref{GAMMA}, four-quark operators containing a pair 
of charm and anti-charm quark fields necessarily arise in the OPE and 
their $B$ meson expectation values do not vanish exactly. These 
effects will be discussed in a forthcoming dedicated publication
\cite{ic}. Here we give a brief introduction and state the salient
conclusions of that analysis.

In discussing such effects in the OPE framework, it is essential 
to distinguish between an expansion in $1/m_b$
(actually, 
$1/(m_b\!-\!m_c)$) and in $1/m_c$. For the sake of clarity we 
resort to a somewhat idealized scenario where the hierarchy
between the two scales is stretched up compared to actual QCD.

To the leading third order in $1/m_b$ there are four such operators of
which two are relevant for $\gsl$:
\bea
\nonumber
H_c=\aver{O_1^c} &\msp{-3}=\msp{-3}& 
-\frac{1}{2M_B}\matel{B}{\bar{b} \gamma_i(1-\gamma_5) c \,
\bar{c}\gamma^i(1-\gamma_5)b}{B}\\
F_c=\aver{O_2^c} &\msp{-3}=\msp{-3}& 
-\frac{1}{2M_B}\matel{B}{\bar{b} \mbox{$\frac{\lambda^a}{2}$}
\gamma_i(1-\gamma_5) c \,
\bar{c}\mbox{$\frac{\lambda^a}{2}$}\gamma^i(1-\gamma_5)b}{B}\;. \qquad
\label{102}
\eea
($F_c$ is the tree-level counterpart of $\tilde H_c$ in Eq.~(\ref{10})).
To higher orders in $1/m_b$ there are additional operators with
derivatives acting on the quark fields. They can be reasonably
neglected in the total width. For general purposes it is convenient to
consider the extended set of four such operators
\bea
\nonumber
O_V^s=\bar{b}b \, \bar{c}\gamma_0c  \msp{7}& & 
O_A^s=\bar{b}\vec\sigma b \, \bar{c}\vec\gamma\gamma_5 c \\
O_V^o=\bar{b}\mbox{$\frac{\lambda^a}{2}$} b \,
\bar{c}\mbox{$\frac{\lambda^a}{2}$} \gamma_0 c & &
O_A^o=\bar{b}\mbox{$\frac{\lambda^a}{2}$} \vec\sigma b \, 
\bar{c}\mbox{$\frac{\lambda^a}{2}$} \vec\gamma\gamma_5 c  \qquad 
\label{104}
\eea
which define the expectation values $H_c$ and $F_c$ by virtue
of Fierz transformations: 
%%% \note{Convention is that $\bar{b}\gamma_0b=+1$}
\bea
\nonumber
O_1^c &\msp{-3}=\msp{-3}&-\frac{3}{2N_c} O_V^s + \frac{1}{2N_c} O_A^s
-3 O_V^o + O_A^o\\
O_2^c &\msp{-3}=\msp{-3}& -\frac{3}{4}(1-\mbox{$\frac{1}{N_c^2}$})
O_V^s 
+ \frac{1}{4}(1-\mbox{$\frac{1}{N_c^2}$}) O_A^s + \frac{3}{2N_c} O_V^o - 
\frac{1}{2N_c}O_A^o
\;.
\label{106}
\eea
The indices $V,A$ and $s,o$ correspond to
products of vector or axial, and color-singlet 
or -octet flavor-diagonal $\bar{b}b$ and $\bar{c}c$ currents.
Below we shall mainly refer to the contributions related to $H_c$ as
most relevant for $\gsl$.

The existence of such an effect 
can be understood since this expectation
value simply states that the propagation of the decay quark
inside $B$ meson is not totally perturbative. It is related to
local expectation values because 
we consider the inclusive width coming from $c$ quark momenta much smaller
than the energy release where the momentum of the lepton pair does not 
differs much from $q_\mu\!=\!(E_r,0,0,0)$, 
and we expand in this difference
\cite{WA}. This inclusive probability for transitions to charm quarks
is affected by nonperturbative dynamics to a lesser degree than for
decays to  light quarks; yet it is 
evident that it should be present at some
level. Its magnitude can be qualified and estimated \cite{ic}. The
relation of these effects to the ``Intrinsic Charm'' in $B$ mesons is
discussed there.

The charm quark propagator undergoes nonperturbative effects even when
it is hard or has large energy or momentum. These effects, however are
accounted for in the standard OPE; to the leading order they are given in
terms of $\mu_\pi^2$ and $\mu_G^2$ as exemplified by the sum rules
considered in Ref.~\cite{optical}. The effect of the Darwin term comes
from both hard and soft charm configurations. Hence, in the Wilsonian
approach a part of the effect of the Darwin operator in Eq.~(\ref{10})
actually represents the contribution $H_c$ of the four-quark charm 
operator. This part belonging to $\rho_D^3$ depends on the choice of
the  normalization 
point $\mu$ and vanishes
if $\mu$ is taken well below $m_c$ -- this option is possible only for
a heavy quark.

If the charm field can be considered as heavy in the scale of typical
hadronic masses $\mu_{\rm hadr}$, one can apply heavy quark expansion 
to these nonperturbative expectation values. 
It proceeds, however, in $1/m_c$ and not in $1/m_b$: 
\beq
\aver{O^c_j}= \sum_k C_{jk}\frac{1}{2M_B}
\frac{\matel{B}{\bar{b}O_k b(0)}{B}}{m_c^{d_k-3}}
\;.
\label{120}
\eeq
Similar to the OPE for inclusive widths, it includes only {\tt local} 
$\bar{b}...b$ operators which, however, now involve only light 
(gluon and $u,d,s$
quark) fields. In the above equation $d_k$ denotes the dimension of the
light field operator $O_k$. Therefore, without introducing the
explicit operators with charm, there exists a chain of higher-order
power terms scaling like $\frac{\Lam^n}{m_b^3m_c^{n-3}}$ rather than 
$\frac{\Lam^n}{m_b^n}$. It is instructive to keep in mind that the large
coefficient of the Darwin operator is associated with the lowest term in this
subseries corresponding to $n\!=\!3$ which gives rise to the 
$\ln{\frac{1}{m_c}}$ enhancement.

The $1/m_c$ expansion of the charmed expectation values thus requires
classifying possible operators $O_k$ and computing the coefficients
$C_{jk}$.\footnote{Alternatively, one can integrate the sum rules for
the corresponding combinations of the zeroth moments (for the
leading-$m_b$ operators) of the heavy
quark structure functions studied in \cite{optical}, over $\vec{q}$. 
This would require, however extending the expansion at least to two
more orders in $1/m_c$.} It is shown \cite{ic} that at the 
tree level there are no
contributions to order $1/m_c$, while $1/m_c^2$ effects are driven by a
few operators, depending on the particular color and Lorentz
structure.\footnote{Say, no contribution for the color-straight
operator with vector currents, one for the product of 
color-straight axial currents, etc.} The dominant in $N_c$ structure
for semileptonic decay is color octet, see Eq.~(\ref{106}), 
for which $1/m_c^2$ effects are
present for both vector and axial currents.

It is interesting, however, that absence of $1/m_c$ nonperturbative
charm loop effects can be vitiated once perturbative corrections are
included \cite{ic}. 
There are two spin-singlet operators $O_{E^2}$ and $O_{E\!\cdot\!E}$ 
and one spin-dependent $O_{E\!\times \!E}$ operator
which emerge already to order $\alpha_s(m_c)$ from the dominant (in
semileptonic decays) color-octet operators, vector and axial,
respectively: 
\bea
O_{E^2} \msp{-4}&=&\msp{-4} \frac{1}{16\pi^2 } \;
\bar{b}\: 
{\rm Tr}\,\vec{E}^{\,2} \:b(0)= 
%%% \msp{-4}&=&\msp{-4} 
\frac{1}{16\pi^2} \;  
\bar{b}\,{\rm Tr}\,(\pi_k
\pi_0 \pi_0 \pi_k) \,b \\ \nonumber
O_{E\!\cdot\!E} \msp{-4}&=&\msp{-4} \frac{1}{16\pi^2 } \;
\bar{b}\: 
(1\!-\!\mbox{$\frac{1}{N_c}$}{\rm Tr}\,)\vec{E}\cdot\vec{E}\:b(0)= 
%%% \msp{-4}&=&\msp{-4} 
\frac{1}{16\pi^2} \;  
\bar{b}\, (1\!-\!\mbox{$\frac{1}{N_c}$}{\rm Tr}\,)(\pi_k
\pi_0 \pi_0 \pi_k) \,b \\ \nonumber
O_{E\!\times \!E}\msp{-4}&=&\msp{-4} \frac{1}{16\pi^2 } \;i\,
\bar{b}\: \vec\sigma \!\cdot\!
\vec{E}\!\times \!\vec{E} \:b(0) 
%%% \msp{-4}&=&\msp{-4} 
= -\frac{1}{16\pi^2} \;i\, \epsilon^{\mu\nu\rho\lambda}\: 
\bar{b}\,\gamma_\mu\gamma_5 \pi_\nu
\pi_\rho \pi_0 \pi_\lambda \,b 
\;,
\label{128}
\eea
where $\vec{E}$ is the non-Abelian chromoelectric field strength (it
includes coupling $g_s$), and
$\pi_\mu$ are the nonrelativistic energy-momentum operators. 
An example of the Lorentz-invariant operator yielding such spin-dependent
interaction is
$$
\frac{1}{16\pi^2 m_b} \; \epsilon^{\mu\nu\rho\lambda}\: 
\bar{b}\,\gamma_\mu\gamma_5 D_\nu D_\rho D_\alpha D_\lambda D^\alpha\,b
$$ 
with $D_\mu$ denoting the full-QCD covariant derivatives. 
Such operators 
come with  coefficient $\propto \!\alpha_s(m_c)/\pi$ 
and are not expected
to dominate for actual charm quark. They would represent the leading
effect in the true heavy charm limit. Physically $1/m_c$
corrections can describe
interference between the (gluon-dressed) parton decay amplitude and
internal $c\bar{c}$ component of the meson, see Fig.~1 
similar to interference 
discussed in detail in Ref.~\cite{mirage}. Soft fields shape the
internal charm component of wavefunction, while short-distance effects
here are characterized by virtuality starting at $m_c$ and governed by
$\alpha_s(m_c)$.
To even higher order in $\alpha_s(m_c)$, and/or including gluons with
momentum scaling as $m_b$ more operators become possible.

\begin{figure}[hhh]
\vspace*{-3mm}
\begin{center}
\mbox{\psfig{file=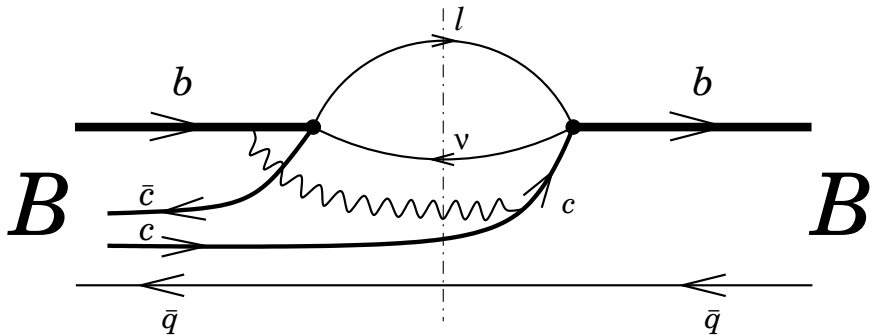,width=12cm}}
\end{center}\vspace*{-10mm}
\caption{{\small 
Example of the interference contribution to the decay width 
suppressed by a single power of $1/m_c$. Shown gluon is hard. All cuts
should be included to maintain the correct scaling.}}
\end{figure}

The effects of intrinsic charm in the context of DIS on nucleons were 
thoroughly investigated in Ref.~\cite{maxim}, though without
perturbative effects. The relevant
charm operators are quark bilinears in this case, yielding quite
different pattern of the power expansion. Including 
hard perturbative corrections can change leading power in $1/m_c$ for
heavy quark decays, as noted above.

The practical OPE \,Eq.~(\ref{120}) in the form of 
an expansion in $\Lam/m_c$ is only asymptotic; the charm expectation values
contain also ``exponential'' pieces like $e^{-2m_c/\mhad}$. 
Various considerations lead us to expect  that for actual charm with 
$m_c\!\simeq\! 1.2\GeV$  these exponential terms can be, if not dominant, at
least not too far suppressed compared to the leading powerlike effects.
These contributions would represent the clearest and unambiguous example of
genuinely independent effects associated with nonperturbative corrections for 
charm quarks in $B$ mesons.

All these contributions, which are suppressed by powers of $1/m_c$ and
therefore 
could be sizeable thus posing a danger to the $1/m_Q$ expansion,
just reflect the 
existence of operators in the OPE of the width containing pairs 
of $c$ and $\bar c$ 
fields, namely $O_j^c$ (and their higher-dimensional analogies). Their 
impact is best analyzed directly through the $B$ expectation values of
$O_j^c$.
The size of these nonperturbative expectation values is
however not very certain. For $\Gamma_{\rm sl}(b\!\to\! c)$ there 
is a single
relevant operator, Eq.~(\ref{24}). The size of its 
expectation value $|H_c|$ is estimated \cite{ic} 
\beq
|H_c|
\lsim 0.005\GeV^3
\label{130}
\eeq 
which translates into a possible
contribution to the width up to $1\:\mbox{to}\:1.5\%$. This looks
reasonable if we compare it with the effect of the Darwin expectation
value itself which, according to our master equations is estimated to
be about $3\%$. A fraction of the Darwin term effect, on the
other hand, represents the leading, $1/m_c^0$ piece of the four-quark
operators with charm.

An alternative way to arrive at a similar numerical estimate is
calibrating 
them with the size of nonfactorizable contribution of four-quark
operators $\bar{b}b \,\bar{q}q$ with light quarks $q$, which would 
enter the
total semileptonic $b\!\to\! u$ width. The four-quark operators with 
non-valence light quark were estimated in
Ref.~\cite{vub} to affect the width at a couple percent level,
employing the information about $D$ meson widths. Assuming that the
expectation values with the charm field replacing light non-valence
quarks exhibit extra suppression due to the charm mass by a factor 
of four,
and accounting for the difference in the phase space in the two
decays, we
end up with an estimate somewhat below the one percent level in 
$\Gamma_{\rm sl}(B)$.

%%%%%%%%%%%%%%%%%
\subsubsection{Experimental constraints}
%%%%%%%%%%%%%%%

A more robust approach is to constrain the charm-related
expectation values directly from experiment. 
Available studies of the
semileptonic distributions already yield some more indirect bounds on such
effects, however the limits fall quite short of the scale expected 
theoretically. One reason behind this insensitivity is transparent:
they manifest themselves in the same way as the soft part of the
Darwin term, and therefore to some extent can be
absorbed by redefining the effective value of
$\rho_D^3$. In particular, for sufficiently inclusive characteristics
replacing
\beq
d\,\rho_D^3 \longrightarrow d \,\rho_D^3 + 32\pi^2 (1\!-\!\sqrt{r})^2
H_c
\label{200}
\eeq
may account for the bulk of the effect, and the actual
difference in the effect of Darwin operator and the charm-related
nonperturbative four-quark expectation value $H_c$ will be revealed at
a suppressed level only in more subtle quantities, say in the
dependence of the hadronic distributions on lepton energy.

There can be various physical manifestations of the nonperturbative
effects associated with moderately short-distance charm quarks in
non-charm beauty hadrons.  A number of effects have been discussed in
connection to the ``intrinsic charm'' hypothesis \cite{brodic}; it is
not always easy to disentangle them from other underlying effects,
however.

Here we are concerned with a more concrete aspect, namely the way such
effects may manifest themselves in the inclusive semileptonic decays
via local $b$-quark operators with charm, Eqs.~(\ref{104}). 
The search for an unambiguous manifestation is complicated by
the expected percent scale in $\gsl$. Yet even any
higher direct upper bound would constitute a very valuable information
complementary to theoretical estimates.

In the heavy quark expansion these four-quark operators would most
directly show up as $\delta$-like contributions near the
quasi--two-body kinematics $b\to\mbox{slow}\:c + \ell\nu$, 
which naively corresponds to such leading terms
in the OPE, with
\bea
\nonumber
\frac{1}{\Gamma_{\rm sl}}\frac{{\rm d}\Gamma_{cc}}{{\rm d}q^2} 
&\msp{-3}\simeq\msp{-3}&
\frac{16\pi^2}{M_B\!-\!M_{D^*}} 
\,\frac{H_c}{m_b^3}\, \delta\left(\sqrt{q^2}\!-\!(M_B\!-\!M_{D^*})\right)\,,\\
\nonumber
\frac{1}{\Gamma_{\rm sl}}\frac{{\rm d}\Gamma_{cc}}{{\rm d}E_X} 
&\msp{-3}\simeq\msp{-3}&
32\pi^2 \,\frac{H_c}{m_b^3}\, 
\delta\left(E_X\!-\!(M_B\!-\!M_{D^*})\right)\,,\\
\frac{1}{\Gamma_{\rm sl}}\frac{{\rm d}\Gamma_{cc}}{{\rm d}E_\ell} 
&\msp{-3}\simeq\msp{-3}&
32\pi^2 \,\frac{H_c}{m_b^3}\, 
\delta\left(E_\ell \!-\!\mbox{$\frac{M_B\!-\!M_{D^*}}{2}$}
\right)\,,
\label{202}
\eea
etc. Placed in the narrow slice of the whole kinematic domain,
these effects are locally enhanced and potentially detectable in
differential distributions, much in the same way as was suggested for
nonfactorizable effects in $\bar{b}b\,\bar{q}q$-type operators with
light quarks in the lepton spectra \cite{WA}, or $M_X^2$ or $E_X$
distributions \cite{distmx}.

\begin{figure}[hhh]
\vspace*{-7mm}
\begin{center}
\mbox{\psfig{file=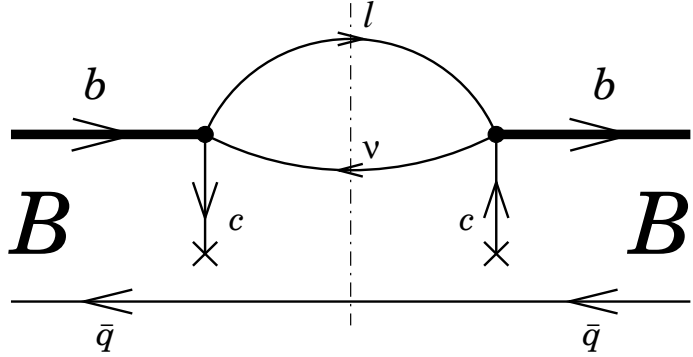,width=10cm}}
\end{center}\vspace*{-12mm}
\caption{{\small 
Nonperturbative charm-field effects in the OPE for the inclusive
decays. With charm fields being low-momentum, the kinematics is
close to two-body for leptons, with $\,\sqrt{q^2}\!\simeq\!m_b\!-\!m_c$\,.
}}\vspace*{-.5mm}
\end{figure}

There are also similar operators with derivatives. In the total widths
they are suppressed by additional
powers of $1/m_b$. They lead to smearing the above quasi--two-body
peaks over the domain of typical width $\sim \!\Lam$. The smearing in
practice can stretch over a sizable  fraction of the allowed kinematic
range for actual $m_b$.\footnote{Perturbative corrections associated
with the anomalous dimensions of the leading operators yield slower
decreasing tails as discussed in Ref.~\cite{WA}; yet in practice this
perturbative 
smearing should hardly be relevant.} It should be noted that the charm
operators can lead to both enhancement or depletion of the decay rates,
depending on the sign of the expectation values.

The above pattern suggests the strategy of studying the integral over the
particular low-$E_X$ kinematic domain and comparing it with the
expectations based on the parameters extracted from the total
moments. It turns out that the
overall integrated hadronic energy
$\aver{E_X}$ and the invariant mass $\aver{M_X^2}$ are theoretically 
expected to depend on practically one and the same combination of
parameters 
$m_b\!-\!0.67m_c+0.1\mu_\pi^2\!-\!0.25\rho_D^3$. Comparing the two
averages experimentally would then be a cross check of possible additional
effects, including charm expectation values.

It seems to
us that the available accuracy in the fully integrated
moments (with respect to $E_\ell$, $M_X$ or $q^2$) does not allow at
present to constrain the expectation values, in particular $H_c$ with
the precision even nearly matching the theoretical expectations
Eq.~(\ref{130}). An interesting possibility is to study the partially
integrated over $E_X$ double differential distribution
\beq
\Delta(\vec{q}^{\,2}; M)=\frac{1}{\Gamma_{\rm sl}}\,
\int_{M_D^2}^{M_{D^*}+M} {\rm d}M_X^2 \;
\frac{{\rm d}^2\Gamma_{\rm sl}}{{\rm d}M_X^2{\rm d}\vec{q}^{\:2}}
\;\,.
\label{210}
\eeq
For the four-fermion operators it is expected to exhibit a more or
less flat contribution above a hadronic scale $M_0$ for moderate
$\vec{q}^{\,2}$, 
$$
\int {\rm d}\vec{q}^{\,2}\: \Delta(\vec{q}^{\,2}) 
\simeq 32\pi^2\: \frac{H_c}{m_b^3}
$$
which however fades away with increase of $\vec{q}^{\,2}$. In this 
scenario the effect can
originate from a limited range of $M_X$
below a hadronic mass $M_{D^*}+M_0$ with $M_0$ expected to be 
around $m_c$ and/or $\mu_{\rm hadr}\approx 1\GeV$.

The most precision data available now are from CLEO lepton moments
measured with a lower cut $1.5\GeV$ \cite{CLEOlm}. However,
the average lepton energy in this interval is $\aver{E_\ell}\simeq
1.73\GeV$ which is very close to the expected location of the
discussed contributions
\beq
E_\ell^{cc}\approx \frac{M_B\!-\!M_D^*}{2} \simeq 1.64\GeV\;.
\label{214}
\eeq
This strongly suppresses the sensitivity of these moments to $H_c$. At
the same time, this physical example illustrates the fact that 
the integrated widths
with cuts are far more vulnerable to higher-order effects: the
relative sensitivity of just the rate with $E_\ell > 1.5\GeV$ to $H_c$ can
naturally be three times higher than of $\Gamma_{\rm sl}^{\rm
tot}(b\!\to\!c)$ simply because such a partial width comprises only
about $30\%$ of the total rate, and impact on the CLEO's $R_0$ 
can be even more
significant. Yet we think it may be premature to attribute some
apparent discrepancy between, say, measured values of $R_0$ and $R_1$ to
effects of charm operators, in view of insufficient control over usual
higher-dimension operators for the widths with such a cut \cite{amst}.

It is also worth emphasizing that at a high enough cut on $E_\ell$ the
operators describing the addressed nonperturbative effects cease to
remain local (in space). Whether the cut at $1.5\GeV$ is safe in this
respect, must yet be understood.

%%%%%%%%%%%%%%%%%%%%%%%%%
\subsection{Heavy quark operators of {\boldmath $\,D\!\ge\! 7$}}
%%%%%%%%%%%%%%%%%%
\label{3.4}

The important question about the achievable accuracy of the OPE for
$\Gamma_{\rm sl}(b\!\to\! c)$ is the magnitude of higher-order power
corrections beyond $1/m_b^3$ terms. Their number proliferates, and it
might look hopeless to get a meaningful answer. Nevertheless, we shall
argue that the effect of the higher-dimensional operators while applying
the proper Wilsonian procedure should be suppressed and not exceed the 
percent level. One also needs to qualitatively  understand the 
hierarchy of the computed corrections through order $1/m_b^3$. This is
discussed later in this section.

To get the way around apparent complexity of higher-order corrections,
we use a few basic ideas.
A few different physical 
momentum scales enter the problem even if one abstracts from 
perturbative corrections generally messing up physics at different
distances. Namely, a number of corrections are driven solely by the
scale $1/2m_b$ -- among those are preasymptotic corrections to the
expectation values of nonperturbative operators over the actual $B$
meson state. It is clear that these terms among higher-order
operators can be neglected at once.

The second largest scale specific to the inclusive width is the energy 
release $m_b\!-\!m_c\simeq 3.5\GeV$. Since
it still is significant, the higher-dimension operators controlled
by such an expansion parameter yield the contribution far below the percent
level as is illustrated by the $LS$ term, unless their effect is
particularly enhanced. There are two reasons for a possible
enhancement of the coefficient functions. One is the ``infrared
instability'' when $m_c/m_b\to 0$. It was discussed in
Sect.~\ref{ic}, and they are taken care of by introducing the explicit
expectation values $H_c$, $\tilde H_c$ of the four-quark operators with charm 
in Eq.~(\ref{10}). Indeed, infrared sensitivity is possible only for
soft $c$ quark line, with the lepton pair carrying the momentum close
to $m_b\!-\!m_c$. This effectively contracts the two weak
vertices in space and time yielding local  four-fermion operator
$\bar{b}\Gamma c\,\bar{c}\Gamma b$.

The second reason discussed in detail in Ref.~\cite{five} is due 
to presence of the large 
parameter $N\!=\!5$ describing the power $N$ with which
heavy quark masses enter the total width, $\Gamma\propto (m_b,
m_b\!-\!m_c)^5$. Therefore, to find the way through the mace of 
higher-order power corrections, we will use the classification of the
coefficients over this parameter.\footnote{The increase in the Wilson
coefficients should be distinguished from the expected factorial growth
of the matrix elements of the operators compared to the dimensional
estimate $\mu_{\rm hadr}^k$; the latter is related to violation of
local duality \cite{shiftasi,inst} and will be addressed 
separately.} Namely, we can require an extra power of the 
hadronic scale $\mu_{\rm hadr}/m_b$ in the
nonperturbative corrections to be accompanied by the maximal power of $N$.

The $1/N$ classification is quite transparent and yields a number of
useful facts. As pointed out in Ref.~\cite{five}, large $N$ pushes one
to the so-called Extended SV regime, since large $N$ places most of
the weight on maximal $q^2$ of the lepton pair. It is also evident
that this dominance is further enhanced for higher-dimensional
operators: the leading contribution comes from multiple
differentiation of the $c$ quark propagator with respect to $m_c^2$, 
which pushes it further and
further into the infrared. In this respect the effect is the same as
for the first mechanism, infrared enhancement, and in the same way the
largest pieces are absorbed into the explicit charmed expectation
values. The upshot of this approach is two-fold:

$\bullet$ The dominant operators are spin-singlet like the kinetic and
Darwin terms, as opposed to the spin-dependent chromomagnetic or $LS$
operators. These are among those which determine higher moments of the
light-cone distribution function. The light-cone set of operators
would dominate if $m_b$ were really much greater than $m_c$ so that
$m_b/N>m_c$, the case  where normalization 
point $\mu$ for the four-fermion
operators with charm could be taken well above $m_c$.

$\bullet$ In the realistic case, the leading ones
among these spin-singlet operators are those
containing more time derivatives; extra space derivatives generally
yield subleading in $N$ effects. Since the left-most and right-most
covariant derivatives in $\aver{\bar{b}D...D b}$ must be space-like
for the heavy $b$ quark, it is actually the SV-type operators 
determining the SV heavy quark distribution function that dominate the
higher-order power corrections. This is in a clear accord with the
onset of the Extended SV regime mentioned above.

It is worth noting that the dominance of the SV-like operators makes
the estimates more conclusive. They can be fully defined with
explicit normalization-point dependence in the same way as the
operators through $D\!=\!6$ , see, e.g.\, \cite{dipole,ioffe,chrom}:
\beq
\aver{\bar{b}\,iD_k(iD_0)^n iD_k \,b}_\mu=3\sum_{\epsilon\!<\!\mu}
\left[2(\epsilon^{(l)}_{3/2})^{n+2}|\tau_{3/2}^{(l)}|^2 + 
(\epsilon^{(m)}_{1/2})^{n+2}|\tau_{1/2}^{(m)}|^2\right] \;.
\label{160}
\eeq
This eliminates the notorious problem with fixing the ad hoc factors which
often can be arbitrarily reshuffled between the operators and
their Wilson coefficients, obscuring numerical estimates. 
For instance, 
the parameter $\lambda_2$ used in some papers
historically was defined with the factor $1/3$ of the actual size of the
physical chromomagnetic operator in $B$, and  
HQET's  $\lambda_1$ 
was laxly defined as minus the expectation value of the
positive operator.

The chain of SV expectation values has a
definite sign. In addition, their scaling 
is given by a physical mass, the energy
of the excited $P$-wave states at which one observes the onset of the
perturbative regime. The recent data on the $M_X$ spectrum in
semileptonic $B$ decays \cite{babar,DELPHIMX} suggest that this scale
is near the expected $1\GeV$.
It is clear that the magnitude of the ratio of
this mass to the energy release $m_b\!-\!m_c$ would govern the rate of
convergence of power expansion. For the SV regime, for example, this is 
explicit in the analysis of the Orsay group \cite{orsaySV}. The
detailed study of the $M_X$ spectrum in semileptonic decays will
allow even some more quantitative bounds on the leading higher-order
heavy quark operators.

The analysis of possible higher-order effects conducted along these
lines suggests that the effect of higher-order operators with
$D\!\ge\!7$ constitutes at most a fraction of percent in 
$\Gamma_{\rm sl}(b\!\to\!c)$, at least if 
the four-quark operators with charm field
are properly introduced to take care of the infrared contributions. It
is worth commenting in this respect 
on the numerical pattern in low-order power
corrections, which is not always properly interpreted and may lead to
unjustified suspicions about convergence.

According to Eq.~(\ref{SLwid}), the first few power corrections
amount, in fractional units to 
\beq
\delta_{\mu_\pi^2}\simeq -0.012\,,\;\;\;\delta_{\mu_G^2}
\simeq -0.035\,,\;\;\;
\delta_{\rho_D^3}\simeq -0.035\,,\;\;\;
\delta_{\rho_{LS}^3}\simeq -0.003\,.
\label{168}
\eeq
The values $\mu_\pi^2\!=\!0.4\GeV^2$, $\mu_G^2\!=\!0.35\GeV^2$, 
$\rho_D^3\!=\!0.2\GeV^3$ and $\rho_{LS}^3\!=\!-0.15\GeV^3$ have been
used. The significant contribution of the Darwin term compared to the
second-order corrections, and in particular, to the kinetic operator,
may raise some concern. In fact, applying the above theoretical 
perspective one can realize that the Darwin operator has the
expected, normal size.

First of all, as detailed earlier, a significant part of the Darwin 
expectation
value actually represents the contribution of the four-quark charm
operator $H_c$ in Eq.~(\ref{24}) if it is understood in the Wilsonian
sense normalized at the appropriate scale, say at $\mu\!\simeq\!
m_c\!+\!1\GeV$. The remaining part is then noticeably smaller
than the estimate of Eq.~(\ref{168}).

It is more important to appreciate the following fact: the effect of the
kinetic and chromomagnetic operators is actually essentially
suppressed compared to a priori expectations based on the scale
arguments. These are nontrivial dynamic results of the OPE which
cannot be accounted for by simple dimensional considerations.

The most obvious example comes from $1/m_b$ effects. Dimensional
estimates would yield 
$$
\delta_{1/m_b} \simeq 5\frac{\La}{m_b}\,,
$$
the result typically exhibited by naive quark model calculations. 
Such a correction would actually belong to the
SV family in Eq.~(\ref{160}): according to Voloshin's ``optical'' sum
rule $\frac{3}{2}\La$ amounts to the expectation value with
$n\!=\!-1$. Such corrections physically exits. However, the OPE ensures
that in QCD they exactly cancel between the binding effects in
the initial $B$ state and hadronization corrections in the final
state. Clearly one would not interpret the fact that $1/m_b^2$
corrections to total widths are infinitely larger than 
the absent $1/m_b$ effects, as a non-convergence of the power series!

It is even more interesting that a similar cancellation, in a sense
holds even to second order in $1/m_b$. Naively one would expect
the kinetic operator to appear with the coefficient scaling like
$5\!\cdot\! 4 / 2\gg 1$, and to have powers of $m_b\!-\!m_c$ rather than
$m_b$ in denominator. However,
the OPE once again ensures that such an effect is totally absent from
the inclusive width if strong interactions are described by a gauge
theory like QCD.\footnote{Even refined parton-based models  
typically predict a
large positive effect of the kinetic expectation value; see, for
example the light-cone based approach of Ref.~\cite{jin} yielding the
coefficient $+35/6$.} The only effect of the kinetic operator emerges as
the $1/m_b^2$ correction to the expectation value 
$\matel{B}{\bar{b} b}{B}$ \cite{buv,motion,optical}. This makes evident
why its effect is numerically suppressed. We can safely discard
such contributions for higher-order operators, which has been already
assumed in formulating the rules of the analysis earlier in this Section.

The coefficient of the chromomagnetic operator does not vanish. Yet it
is still partially suppressed, which can be explicitly seen computing
its effect when employing $N\!=\!5$ as a free parameter: it
scales like $N$, but not $N^2$. This is related to the
spin-nonsinglet structure of the chromomagnetic operator -- it does
not belong to the SV family of operators of Eq.~(\ref{160}) being an
antisymmetric combination of the spacelike derivatives. This partial
suppression is also
reflected in the dependence of its coefficient in the width: compared
to the parton width it is
enhanced by a single power of $m_b/(m_b\!-\!m_c)$.

A similar partial suppression is present for all higher-order
spin-nonsinglet operators, which is precisely illustrated by the $LS$
contribution. It has a coefficient scaling like that of $\mu_G^2$,
and its numerical contribution is safely at a sub-percent level.

For the Darwin operator, however, the cancellations `accidentally'
enforced on the lower-dimension SV parameters in the OPE 
do not hold anymore, and
its coefficient has `normal' or expected magnitude. Its numerical
impact on the total width is on general grounds expected to be in a 
few percent range,
assuming the normal scaling of the expectation values following from
the family of the SV sum rules \cite{rev,ioffe}. The effect of
higher-order operators then should not exceed the scale of a
half percent.

To summarize, we expect, and have argued that the contribution of 
heavy quark operators with $D\!\ge\!7$ can give only a sub-\% effect
on $\gsl$ provided the four-fermion
operators with charm field in the Wilsonian OPE have been properly 
incorporated.

%%%%%%%%%%%%
\section{Perturbative Contributions}
\label{PERTCONT}
%%%%%%%%%%%

Once the perturbative effects in the quark masses and higher-dimensional
operators have been  properly incorporated, one 
can turn to the radiative corrections to the Wilson coefficients. In
practice, high accuracy is required for the leading (parton) operator
coefficient denoted as $1+A_3^{\rm pert}$ in Eq.~(\ref{10}), and we
routinely refer to it as the perturbative correction $A^{\rm pert}$,
see Eq.~(\ref{SLwid}).

It has become customary to express perturbative corrections as series in the 
$\overline{\rm MS}$ coupling evaluated at $m_b$. Yet the scales typical 
for the radiative corrections are notably lower than $m_b$. 
By evaluating the lowest order term in $\alpha_s^{\overline{\rm MS}}(m_b)$ 
one generally underestimates $A^{\rm pert}$ and 
forces the $\alpha_s^2$ and higher terms to possess  
large coefficients. If, however, at least the second 
order BLM contributions are known, then one can express 
the findings in terms of $\alpha_s^{\overline{\rm MS}}(m_b)$ without 
raising numerical havoc.

It should be noted that using the Wilsonian prescription with an
appropriate hard factorization scale $\mu$ significantly improves
convergence of the perturbative series and makes the above problem less
acute. This is illustrated by Fig.~3 showing the contribution of 
different momentum scales in the total width, with and without
separation of short- and long-distance effects.\footnote{We show the
distribution from Ref.~\cite{blmvcb} which did not incorporate $1/m_b^3$ 
subtraction corresponding to Darwin operator.} The effect of removing
the infrared domain is self-manifest.
Moreover, it is evident that a too significant part of the perturbative
corrections with pole masses comes from gluon momenta below $500\MeV$,
which casts serious doubts on reliability of the numerical results in
this scheme. 

\begin{figure}[hhh]
\begin{center}
\mbox{\psfig{file=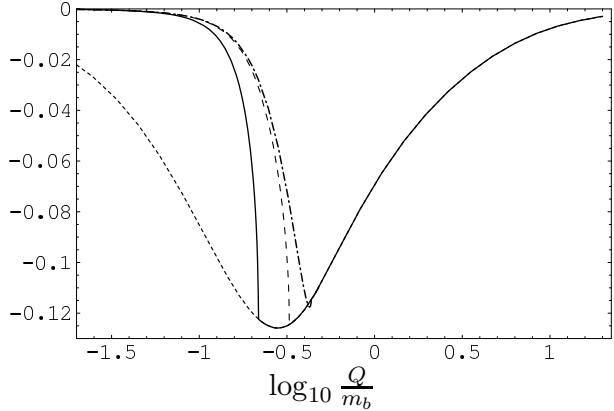,width=8cm}}
\end{center}\vspace*{-5mm}
\caption{{\small 
Gluon momentum scale distribution in 
$\Gamma_{\rm sl}(b\!\to\!c)$. Solid, dashed and dot-dashed lines
correspond to $\mu\!=\!1\GeV$, $\,1.5\GeV\,$ and $\,2\GeV$,
respectively; lighter short-dashed line illustrates the case of 
$\mu\!=\!0$ (pole masses). The area under each curve gives 
the first-order perturbative coefficient.}}
\end{figure}

%%%%%%%%%%
\subsubsection{BLM summation}
\label{BLM}
%%%%%%%%%%

Computing the leading-order ${\cal O}(\alpha_s)$ 
contribution does not tell us at which 
scale the coupling has to be evaluated; that is fixed only by 
terms of higher order in $\alpha_s$. The (extended) BLM prescription 
\cite{BLM,bbbtau,dmw}  is based 
on the conjecture that the bulk of higher order terms can be 
incorporated by replacing the fixed lowest order $\alpha_s$ by the 
scale-dependent $\alpha_s(k^2)$ where $k$ denotes the momentum flowing 
through the gluon line, cf. Fig.~1. The running of $\alpha_s$ is 
determined by the first
coefficient of the Gell-Mann-Low beta function: 
$\beta_0 = \frac{11}{3}N_c - \frac{2}{3}n_f\simeq 9$. The resulting expansion 
is of the type $\alpha_s(1+ \sum _{n=1}^{\infty} c_n(\beta _0\alpha_s)^n)$ 
with $n=1,2,...$ representing the second-order, third- etc.\ BLM 
correction. The BLM prescription 
%%% \note{Why the note that these pieces are clearly present, was removed?}
amounts to the assumption that the perturbative terms left out are 
much smaller since not enhanced by the large value of $\beta_0$.
A dedicated recent discussion can be found in Ref.~\cite{newblm}.

This ansatz has turned out to be intriguingly successful in several 
cases. Often it is used in a simplified form by considering only the
first nontrivial BLM correction -- this constitutes the essence of the
BLM {\it scale fixing}\, for $\alpha_s$ \cite{BLM}.
However in applying this properly to semileptonic 
$B$ decays one has to be aware of some specific complexities: 
\begin{itemize}
\item 
There are effectively two `running' couplings in the problem, 
namely $\alpha_s$ and $m_b$. There is no reason why they should 
be controlled by the same scale -- on the contrary, one expects them 
to be different. Moreover, the standard BLM scale fixing procedure is not
defined -- and rather often becomes meaningless -- in the case of more
than one coupling. Indeed, varying the normalization scale for masses the
BLM scale for $\alpha_s$ varies in the whole range from zero to infinity.
\item 
As stated before good judgment has to be exercised in choosing the 
proper mass construction. The pole mass with its intrinsic infrared 
uncertainty $\sim {\cal O}(\Lam)$ is ill-suited in this
context. Using the BLM improvement only aggravates this problem. 
\end{itemize}

The analytic expressions for the perturbative corrections and the way to
implement the Wilsonian cutoff are described in Appendices \ref{ap2}
and \ref{ap3}. They also give numerical values.

%%%%%%%%%%%%
\subsubsection{Non-BLM contributions}
\label{NONBLM}
%%%%%%%%%%%

The {\tt non}-BLM terms are expected to be small. They
are known in the analytic form neglecting the charm mass (the case of
$\Gamma_{\rm sl}(b\!\to\!u)$) \cite{timo}. In $b\!\to\!c$ a numerical
evaluation exists. The authors of Ref.~\cite{czarm} have computed ${\cal
O}(\alpha_s^2)$  corrections to $b\to c\,\ell\nu$  at the three different 
kinematical points $q^2 = 0$, $m_c^2$ and $(m_b - m_c)^2$, where 
$\sqrt{q^2}$ denotes the mass of the lepton pair. From these three values
one interpolates  over the full $q^2$ range. The non-BLM corrections
indeed turn out to be small for the low-scale masses, which is a
highly welcomed result: the consistent procedure to fix the scale of
$\alpha_s$ in the non-BLM terms is not known unless third-order
corrections beyond simple BLM are available. This would yield
significant numerical uncertainty if the non-BLM coefficients were 
not suppressed.

%%%%%%%%%%%%
\subsection{Overall perturbative correction}
\label{totalpert}
%%%%%%%%%%%

It is gratifying that the perturbative corrections to
$\Gamma_{\rm sl}(b\!\to\!c)$ with the short-distance low-scale masses 
show good convergence \cite{upset}. Let us define
\beq
z_0(r)\,A^{\rm pert}(r;\mu)=z_0(r)+a_1\mbox{$\frac{\alpha_s}{\pi}$}+ 
a_2\left(\mbox{$\frac{\alpha_s}{\pi}$}\right)^2 +
a_3\left(\mbox{$\frac{\alpha_s}{\pi}$}\right)^3 + ...
\label{80}
\eeq
assuming the standard choice of the $\overline{{\rm MS}}$ 
coupling $\alpha_s(M)$
normalized at $M\!=\!m_b$. 
As an example, at $\mu\!=\!1\GeV$, $m_b\!=\!4.6\GeV$ 
and $m_c\!=\!1.15\GeV$ we have
\bea 
\nonumber
a_1\msp{2}&\msp{-5}\simeq \msp{-5} & -0.94\,z_0(r), \qquad  \msp{.2}
a_2^{\rm non-BLM}\simeq -1.0\,z_0(r),  \\ 
a_2^{\rm BLM}&\msp{-5}\simeq \msp{-5} & 
-0.45\beta_0\,z_0(r), \msp{3}\qquad a_3^{\rm BLM}\simeq
-0.21\beta_0^2\,z_0(r),\qquad a_4^{\rm BLM}\simeq +0.09\beta_0^3\,z_0(r),...
\;
\label{82}
\eea
This can be compared with the expansion in terms of pole masses: 
\bea
\tilde a_1\msp{2}&\msp{-5}\simeq \msp{-5} & -1.78\,z_0(r),  \qquad  \msp{.2}
\tilde a_2^{\rm non-BLM}\simeq 1.4\,z_0(r),  \\
 \tilde a_2^{\rm BLM}& \msp{-5}\simeq \msp{-5} &  -1.92\beta_0\,z_0(r),
 \msp{3}\qquad \tilde a_3^{\rm BLM}\simeq
-2.8\beta_0^2\,z_0(r), 
\qquad \tilde a_4^{\rm BLM}\simeq -4.9\beta_0^3\,z_0(r), ...
\; 
\label{84}
\eea
(the second-order BLM coefficient in the pole mass scheme was first 
evaluated in Refs.~\cite{lsw} and its size interpreted as
uncontrollable behavior of the perturbative series for the
semileptonic widths). 
With this improvement, it fortunately does not make a too significant
difference which approximation beyond the two-loop result to adopt, as
illustrated below. Using the complete BLM-resummed result has the
advantage of being nearly independent on the initial scale used to
normalize $\alpha_s$. To this we add the second-order non-BLM
correction evaluated with $\alpha_s\!=\!0.25$ and regard this as our 
central estimate:\footnote{Since the infrared domain is removed to all
orders in perturbation theory, the effective scale cannot be too low.}
\beq
A^{\rm pert}(r;\mu)\simeq 0.908\, \;\; 
\mbox{ at $\;\frac{\mu}{m_b}\!=\! \frac{1}{4.6}\;$ 
and $\;r\!=\!0.0625$}\; .
\label{86}
\eeq
Alternatively, if we keep only the full second-order result evaluated
with $\alpha_s(m_b)=0.22$, this number
becomes $0.909$; including additionally third, or third and fourth BLM
terms yields $A^{\rm pert}(r;\mu)\simeq 0.903$ and $A^{\rm
pert}(r;\mu)\simeq 0.904$, respectively. 

There are a few sources of possible theoretical uncertainties
here. Before addressing them, we should emphasize that they shall {\tt
not} include uncertainties in the numerical values of $\alpha_s(m_b)$
or in the running quark masses -- those are determined from experiment
and in principle have definite values as soon as the scheme is fixed,
although they carry some error bars in practice.

The uncertainty in the second-order non-BLM coefficient even estimated
conservatively \cite{czarm} leads to $\delta A^{\rm pert}$ 
about $0.0025$; we may think it
was actually overestimated. A potentially more significant uncertainty could
come from varying the effective scale in $\alpha_s$ used to evaluate this
term. However, since the magnitude of the non-BLM coefficient in 
Eq.~(\ref{82}) is about unity, this still would not exceed $0.003$.

The real question is then the size of the third- and even higher-order
perturbative corrections. They are expected to be dominated by the BLM
corrections which has been completely accounted for, and yield a
sub-\% correction. Hence, we have reasons to consider the estimate of
the theoretical uncertainty in knowledge of the perturbative factor
$\delta A^{\rm pert}(r; 1\GeV)\!\lsim\! 0.009\;$ as conservative and
justified: 
\beq
A^{\rm pert}(0.25^2; 1\GeV)=0.908 \pm 0.009 \qquad \mbox{ at ~} 
\alpha_s(m_b)\!=\!0.22\;.
\label{90}
\eeq
The dependence on the actual value of $\alpha_s(m_b)$ follows from
Eq.~(\ref{vcb2}). The dependence on the values of heavy quark masses and 
the used normalization scale $\mu$ can be obtained by explicit
evaluation of equations in Appendices \ref{ap2} and \ref{ap3}.

%%%%%%%%%%%%%%%%%%%%%%
\subsection{Wilson coefficients of power-suppressed operators}
%%%%%%%%%%%%%

While the Wilson coefficient of the kinetic operator is identical to
the coefficient of the unit operator $A^{\rm pert}$ and therefore 
known completely to two loops, the $\alpha_s$ corrections to the
chromomagnetic operator have not been calculated so far. Once known, 
they would tell us the corrections to the coefficient for
the $LS$ operator as well. The coefficient for the Darwin operator is
an independent one, and its perturbative corrections are not known
either. Part of the perturbative corrections comes from the
short-distance renormalization of the heavy quark vector and axial
current, and this one always enters as an overall factor. Therefore,
it is reasonable to factor out $A^{\rm pert}$ in the expression for
the width in the absence of explicit calculations of the remaining
corrections.

The accuracy of the tree level value of the exact chromomagnetic 
coefficient $1+A_5^{\rm pert}$ 
is conservatively estimated as $30\%$. It is generally expected
to deteriorate for higher operators, and we assume only a $50\%$
accuracy for the Darwin coefficient; since the latter is still poorly
known and is rather constrained from above by data 
\cite{DELPHI}, this does not introduce a significant additional
uncertainty. The chromomagnetic expectation value, on the contrary is
known rather accurately, Eq.~(\ref{60}). 
The actual normalization scale for total semileptonic width is
probably somewhat higher, which slightly decreases the expected
effective value of $\mu_G^2$ in Eq.~(\ref{SLwid}). We therefore estimate
\beq
\frac{\delta_{A_5} \Gamma_{\rm sl}}{\Gamma_{\rm sl}}\simeq 0.01\;,
\qquad
\frac{\delta_{D} \Gamma_{\rm sl}}{\Gamma_{\rm sl}}\lsim 0.015\;.
\label{98}
\eeq
Lack of evaluation of the ${\cal O}(\alpha_s)$ corrections to
chromomagnetic and Darwin Wilson coefficients becomes now one of the
limiting factors in the overall theoretical precision \cite{amst}.

%%%%%%%%%%%%%%%%
\section{Theoretical Uncertainties in 
\boldmath $\Gamma (B\to X_c\,\ell \nu)$} 
\label{UNCERT}
%%%%%%%%%%%%%%%%%

%%%%%%%%%%%%%%%%%%
\subsection{First yet least: local duality violation}
%%%%%%%%%%%%%%%%

A separate, more conceptual factor limiting the achievable theoretical
accuracy would be violation of local duality. While it is related to 
high orders in the OPE, it has a different 
origin from what was discussed in Sect.~\ref{3.4} -- a factorial
growth of the expectation values of the higher-dimensional operators in
the OPE, which makes the practical OPE series only asymptotic 
\cite{shiftasi,inst}. Both the 
qualitative pattern and numerical aspects of duality violation depend
crucially on the observable in question.

In Ref.~\cite{vadem} a detailed and comprehensive analysis of 
limitations to local duality in $B \!\to \!X_c \,\ell\nu $ was given. Here
we reiterate some of the salient points: 
\begin{itemize}
\item Meaningful statements about duality violation require  
an accurate
definition of the notion. That has been given for the OPE-treatable
quantities like inclusive decay widths, and must be understood as
deviations from the {\sf nonperturbatively corrected} predictions through a
sufficiently high order in the OPE (as opposed to parton-level
estimates). 
\item 
Claims of duality violations one can find in the literature often 
turn out to constitute violations of the OPE. No example of the
latter has been given so far, and actually are not expected to
exist. 
\item 
On quite general grounds one infers that duality violations 
are given by `oscillating' functions {\it a l\'{a}}\, 
$\sin{\frac{m_Q}{\mu_{\rm hadr}}}$\, that get further 
suppressed by powers of the energy release. No systematic excess or
depletion of the rate compared to the OPE is allowed.
\item 
The fully integrated width $\Gamma_{\rm sl}(B)$ relegates local
duality violation {\it per se} \,to rather high orders. I.e., 
limitations of local duality are suppressed by a high power 
of the energy release. However, a much larger impact can exist on 
differential distributions or various widths with cuts.
\item 
It has been estimated that limitations to duality in $\Gamma_{\rm sl}(B)$ 
amount to at most a few permill. This bound cannot be
rigorously proved from QCD without some quantitative understanding of 
details of dynamics governing the nonperturbative regime of
QCD. However, in any scenario effects of local duality violation in
total semileptonic widths are much smaller than those in the
differential distributions or moments used to extract the heavy quark
parameters, thus introducing only negligible corrections in practice
compared to the latter. 
\end{itemize}

\subsection{Summary}

We have expressed the total semileptonic width of $B$ mesons as a 
functions of heavy quark parameters, namely quark masses and 
$B$ expectation values of local operators through order $1/m_Q^3$, 
in addition to $|V_{cb}|^2$. The relation still has a
potential theoretical error due to perturbative uncertainties in the Wilson 
coefficient of the leading $\bar bb$ operator,  the QCD radiative 
corrections in the Wilson coefficients of chromomagnetic 
and Darwin operators, in the impact of `intrinsic charm' and in the
contributions from higher-order  power suppressed contributions. As 
already stated in Eq.(\ref{sumunc}), we estimate their impact 
on $|V_{cb}|$ not to exceed $\pm 0.005$, $\pm 0.012$, $\pm 0.007$ and 
$0.004$, respectively. We think that these estimates are far from 
aggressive; the actual uncertainties might even be smaller. For
instance, as mentioned in Sect.~\ref{NUMBERS}, 
there are correlations between $\alpha_s$ and the value extracted for 
$m_Q$.  
Furthermore, a large part of the contribution 
from the charm operators is accounted for by the Darwin contribution, 
when the latter is extracted from experiment, cf.\ Eq.~(\ref{200}). 
By the way,
this ``IC'' contribution may explain why the central value of
$\rho_D^3$ as suggested by experiment is smaller than anticipated.

In summary: we estimate that the present theoretical
accuracy in expressing $\Gamma_{\rm sl}(B)$ through the heavy quark
parameters corresponds to an uncertainty in $|V_{cb}|$ of about
$2\%$, Eq.~(\ref{sumunc}).

%%%%%%%%%%%%%%
\section{Determining the Heavy Quark Parameters}
\label{HQP}
%%%%%%%%%%%%

Through order $1/m_Q^3$ we have the following heavy quark parameters: 
\begin{enumerate}
\item 
Quark masses $m_b(\mu )$ and $m_c(\mu )$; 
\item 
The chromomagnetic and kinetic expectation values $\mu_{G}^2 (\mu )$ 
and $\mu_{\pi}^2 (\mu )$, respectively; 
\item 
The Darwin and $LS$ terms $\rho_{D}^3(\mu )$ and $\rho_{LS}^3(\mu )$, 
respectively. 
\end{enumerate}
We have excluded the charm operators represented by $H_c$, which are
treated separately.
We want to determine the values of these HQ parameters in a way that does not 
jeopardize the strong points of the OPE expression 
for the width, Eq.~(\ref{SLwid}), namely an expansion in inverse powers 
of (at worst) $m_b\!-\!m_c$, when only local operators are relevant.

The $b$ quark mass has been extracted from beauty production at threshold
in $e^+e^-$ annihilation by several authors \cite{KUEHN}. Their findings 
are completely consistent within the stated uncertainties 
of about $1.5\%$.
The techniques (and moments) employed in the analysis differ 
somewhat from author to author and the agreement in their findings 
is reassuring. One should keep in mind, though, 
that these determinations not only share their experimental input. 
The underlying approach is the same, as well as a number of general
assumptions.  
The value stated in Eq.~(\ref{48}) could thus be subject to 
some systematic bias. Arguments based on the 
SV sum rules actually suggest that 
$m_b^{kin}(1\GeV)$ lies a bit above $4.6\GeV$, if a relatively low
kinetic expectation value extracted from experiment so far is confirmed.

The values for $m_b$ and $m_c$ can be determined also from the shape 
of the energy spectra in semileptonic (or $b\!\to\!s+\gamma$) 
$\,B$ decays, which is most concisely encoded in their first few
moments. Extractions were attempted based on the formulae of
Ref.~\cite{USA}, their status is still controversial. 
Another  relevant for us area of concern in the 
CLEO analysis 
is that the relation of 
Eq.~(\ref{MBMC}) is imposed as a constraint 
for $m_b\!-\!m_c$ in semileptonic decays. 
Irrespective of that it should be noted
that the central value extracted from $e^+e^- \to \bar bb$ and 
from $B$ decays agree 
well even within only the stated errors reported in the latter. 
Considering that the two processes 
have very different systematics experimentally as well as theoretically 
and should not have a correlation with the possible bias alluded to
above, this is a nontrivial and encouraging result.

Since the early days of the HQE one has often employed the expansion 
given in Eq.~(\ref{MBMC}) that relates the difference 
$m_b\!-\!m_c$ and the spin
averaged charm and beauty meson masses.  
Yet in view of the accuracy in $m_b$  
that has been achieved, the relation (\ref{MBMC}) may 
represents the most vulnerable part of the analysis. It 
suffers from two systematic weaknesses: 
\begin{itemize}
\item 
It brings in 
effectively an expansion in 
$1/m_c$ rather than $1/m_b$ or $1/(m_b\!-\!m_c)$. At the very least it
is obvious that $1/(m_b\!-\!m_c)^2$ and $1/m_c^2$ differ a lot in
magnitude, since a short-distance charm mass enters.
\item 
It involves the 
{\tt nonlocal} correlators $\rho_{\pi\pi}^3$ and $\rho_s^3$
intrinsically unrelated 
to $B$ decays. There are actually complementing 
indications that their sizes are 
particularly large.\footnote{It was independently observed analyzing
the actual hadron mass spectrum \cite{chrom}, in the exactly solvable
't~Hooft model \cite{lebur} and in the pilot lattice study \cite{kronsim}.} 
\end{itemize} 
Accordingly one is ill-advised to impose relation 
(\ref{MBMC}) in the analysis. Instead one should check the
validity of the relation if there is an option of determining  
$m_c$ independently. It is possible that
Eq.~(\ref{MBMC}) holds with reasonable accuracy; 
for eigenvalues of Hamiltonian 
often exhibit a more obedient behavior than formfactors etc. However 
we cannot count much on such good behavior a priori.
Since $\mu_{\pi}^2 \!-\! \mu_G^2 \!\ll\! \mu_{\pi}^2$, 
the expansion 
$$
m_b \!-\! m_c = M_B \!-\! M_D + 
\frac{\mu _{\pi}^2 - \mu_G^2}{2} 
\left( \frac{1}{m_c} - \frac{1}{m_b} \right) - 
$$ 
\beq 
\frac{-\rho _D^3- \rho_{LS}^3 + \bar \rho ^3 
+\rho^3_{\pi G} + \rho_A^3}{4}
\left( \frac{1}{m_c^2} - \frac{1}{m_b^2} \right) 
+ {\cal O}(1/m_Q^3) \; , 
\eeq 
i.e., where the pseudoscalar rather than the spin averaged meson masses 
are used, is expected to be more stable with respect to higher 
order power corrections \cite{chrom}. It requires some care
incorporating the
perturbative corrections to $\mu_G^2$, but is more promising in
reducing the uncontrollable errors.

When the primary goal is to determine $|V_{cb}|$ as precisely as 
possible rather than the HQ parameters, another observation becomes
important.  Almost the same combination of the HQ parameters controls 
the low semileptonic moments and the total semileptonic width. It also
controls the first hadronic moment $\aver{M_X^2}$.
This  combination of 
HQ parameters can thus be determined with higher accuracy than the 
individual HQ parameters -- and the error associated with it becomes 
partially experimental. The theoretical status of evaluations of the
moments is not as advanced as for $\Gamma_{\rm sl}$, therefore the
expressions used for them should be viewed with lower confidence than
what we know about the width itself. Nevertheless, we
shall adopt them literally to illustrate the point.

%%%%%%%%%%%%%%%%
\subsection{Numerical examples}
\label{NUMBERS}
%%%%%%%%%%%%

An important new element is provided by the measurements of 
lepton energy and hadronic mass moments in inclusive semileptonic 
$B$ decays, as pioneered by CLEO and achieved also by DELPHI. Such 
data provide us with novel detailed information on the HQ
parameters. Even though they do not allow to pinpoint the values of
$m_b$ and $m_c$ separately, they turn out to improve significantly the
accuracy and credibility of extracting $|V_{cb}|$. To illustrate this,
we adopt evaluation Eqs.~(\ref{vcb1}) and (\ref{vcb2})
neglecting possible uncertainty in $\alpha_s(m_b)$.

Comparing with the predictions for, say, the first lepton moment
$\aver{E_\ell}$ or first hadronic  moment $\aver{M_X^2}$ expressed in
terms of the very same heavy quark parameters, we arrive at
{\small
\begin{eqnarray}
|V_{cb}|\msp{-3}&=&\msp{-3} V_{cb}^0\, \left\{
1-1.70\GeV^{-1}\,(\aver{E_\ell}\!-\!1.383\GeV)
-0.075 \left( m_c(1\GeV) \!-\!1.15\GeV \right )\qquad \right. \nonumber \\
& & \left.\msp{29}+0.07 \left ( \mu_{\pi}^2 \!-\!0.4\GeV^2 \right )
-0.055 \left (\rho_D^3 \!-\!0.2\GeV^3 \right ) \right. \nonumber \\
& & \msp{29}\left. -\,0.085 \left ( \mu_G^2\!-\!0.35\GeV^2 \right) 
-0.005 \left ( \rho_{LS}^3 \!+\!0.15\GeV^3 \right ) \right\} , 
\label{346}
\end{eqnarray}
\vspace*{-7mm}
\begin{eqnarray}
|V_{cb}|\msp{-3}&=&\msp{-3} V_{cb}^0\, \left\{
1+0.14\GeV^{-2}\,(\aver{M_X^2}\!-\!4.54\GeV^2)
-0.03 \left ( m_c(1\GeV) \!-\!1.15\GeV \right )\qquad  \right.\nonumber \\
& & \msp{29} \left. +0.1 \left ( \mu_{\pi}^2 \!-\!0.4\GeV^2 \right )
+0.1 \left (\rho_D^3 \!-\!0.2\GeV^3 \right ) \right. \nonumber \\
& & \msp{29} \left.-\,0.01 \left ( \mu_G^2\!-\!0.35\GeV^2 \right ) 
+0.006 \left ( \rho_{LS}^3 \!+\!0.15\GeV^3 \right ) \right\} .
\label{347}
\end{eqnarray}
}
Taking for orientation the literal error bars on the two moments
quoted by DELPHI \cite{DELPHI} we would arrive at the uncertainties
due to heavy quark parameters at the level of $2.5\%$ and $1.2\%$,
respectively. To have an independent experimental constraints on
kinetic and Darwin expectation values one can use the second and
third hadronic moments, or their improved version \cite{amst}.

The additional uncertainty would come from the charm expectation value
$H_c$. However, as stated by Eq.~(\ref{200}), in such an approach it is
expected to be milder: extracting in practice the
Darwin expectation value, one partially accounts in it for the effects
of possible contribution of $H_c$.

A similar reduction of sensitivity to $\alpha_s$ looks
probable. According to Fig.~1 the effective scale of heavy quark
masses shaping $\gsl$ is somewhat larger than $1\GeV$; if the same
applies to the lowest moments, there should be a correlation between
the extracted values of the masses and $\alpha_s$ reducing the
dependence on the latter. This can be visualized by adopting a higher
normalization scale for masses $\mu\!\simeq\! 1.5$ or even $2\GeV$ which
reduces the overall perturbative correction.

%%%%%%%%%%%%
\section{Summary and Outlook}
\label{OUT}
%%%%%%%%%%

The  heavy quark expansion allows 
to express $\Gamma (B\!\to\! X_c \,\ell\nu)$ as a series  
in powers of $1/(m_b\!-\!m_c)$ with coefficients that are controlled 
by the $B$ expectation values of local HQ operators; the number of
unknowns here through order $1/m_b^3$ is limited. Duality violations 
for this fully integrated rate have been estimated not to exceed the 
few permill level, i.e. to be irrelevant. Contributions from 
unknown higher order corrections -- perturbative as well as 
nonperturbative -- are also estimated to be near or below 1\%. 
It would require a very major effort to improve on this situation; 
at present we see no need for that, except computing the perturbative
corrections to the chromomagnetic and Darwin operators.

The burning issue is how accurately we can determine the 
values of the HQ parameters from other observables. 
In the past the validity of the heavy quark expansion 
for $\Gamma (B\!\to\!X_c\,\ell\nu)$ 
has been compromised by invoking the relation of Eq.~(\ref{MBMC}): 
$m_b\!-\!m_c$ is effectively given by an expansion in $1/m_c$ with 
the additional drawback of uncontrolled non-local correlators 
contributing.

This weak link in the analysis can be overcome now: measuring the energy 
and mass moments in $B$ decays allows us to obtain values for the 
HQ parameters from expressions in powers of $1/(m_b\!-\!m_c)$, where 
contributions from nonlocal operators are absent.
%%% either absent or suppressed. 
Furthermore the HQ parameters the size of which previously 
could be inferred mainly by theoretical arguments, can be measured -- 
and actually in more than one way. Such redundant extractions are 
a powerful tool in establishing theoretical control. It turns out that 
the low leptonic and first hadronic moments are controlled by almost 
the same combination of 
HQ parameters as $\Gamma (B\!\to\!X_c\,\ell\nu)$ . This means that 
uncertainties previously 
viewed -- correctly -- as theoretical, become more experimental and thus 
can be reduced by better data.  The situation there 
-- while promising and encouraging -- is still fluid requiring 
closer experimental as well as theoretical scrutiny. 
On the theoretical side higher order contributions to 
the 
{\em moments} have to be evaluated with the same care as it has been done in 
this paper for the total semileptonic $B$ width. On the experimental side it 
is essential that the cut in the 
lepton energy be kept as low as possible. For unless the major part 
of the spectrum is measured, limitations to local duality 
can enter through the `back door' to haunt us. Furthermore such cuts 
decrease the hardness of the transition and thus deteriorate the 
convergence and reliability of the expansion. 

Extracting $|V_{cb}|$ with neither the theoretical nor the experimental 
uncertainty exceeding the one percent level is thus within our reach. 
It appears 
unlikely that a higher accuracy should ever become necessary.

Finally -- and this is one of the strengths of the OPE -- the same values of 
the heavy quark parameters $m_b$, $\mu_{\pi}^2$, $\rho_D^3$ etc.\ can be
used when extracting $|V_{ub}|$ from $B \!\to\! X_u \,\ell \nu$ transitions.
The latter, however additionally require evaluation of the properly
defined four-quark expectation values with $b$ and light quarks.
\vspace*{2mm}

{\bf Acknowledgments:} We have benefited from exchanges with
S.~Brodsky, S.~Gardner, J.~K\"uhn and A.~Vainshtein. N.U.\ is also grateful
to M.~Battaglia,  M.~Calvi, P.~Gambino and P.~Roudeau  for many related
discussions. Three of us
(D.B., I.B.\ and N.U.) are grateful for the 
hospitality extended to them by the Institut fur Theoretische Teilchenphysik, 
Universit\"at Karlsruhe, where our collaboration was started. This work has 
been supported in part by the National Science Foundation 
under grant number PHY00-87419, by the DFG Research Group
``Quantenfeldtheorie, Computeralgebra und Montre Carlo Simulationen'',
by the DFG Sonderforschungsbereich SFB/TR9 ``Computational Particle Physics''
 and by the German Ministry for Education and Research BMBF under the
grant number 05HT1VKB1 and the DFG Merkator Program.

\renewcommand{\theequation}{A.\arabic{equation}}
\renewcommand{\thesection}{}
\setcounter{equation}{0}
\setcounter{section}{0}
\section{Appendices}
\renewcommand{\thesubsection}{A.\arabic{subsection}}

\subsection{Heavy quark masses} \label{apmQ}

Having stated in Sect.~\ref{QUARKMASS} our preferences for 
which quark masses to adopt, we mention here how one can translate 
between the different `quark mass languages'.

The usually used $\overline {\rm MS}$ mass 
$m_Q(\mu )$ is not really a quantity in the Lagrangian, it is rather 
a combination of parameters optimized for perturbative 
calculations in dimensional regularization. Its perturbative relation to the 
pole mass
\bea
\nonumber
m^{\rm pole} &\!\!=&\;  \overline{m}(\overline{m}) \left\{1+ \frac{4}{3}
 \frac{\asMS (\overline{m})}{\pi} +\left(\frac{\asMS 
(\overline{m})}{\pi}\right)^2
 \left[\frac{\beta_0}{2}\left(\frac{\pi^2}{6}+\frac{71}{48}\right)
  \right. \right. \\
&& \left. \left.  + \frac{665}{144} + \frac{\pi^2}{18}\left(2 \ln 2
 - \frac{19}{2}\right) - \frac{1}{6} \zeta (3) -
\frac{8}{3}\right]\right\}+...
\label{msmass}
\eea 
is known now to three loops \cite{mass3loop}.
At $\mu \geq m_Q$ the $\overline {\rm MS}$ mass coincides more or less 
with the running Lagrangian mass probed at scale $\propto \!\mu$, and thus
is appropriately used to describe high energy processes, $E\gsim
m_Q$. However maintaining by definition the same $\mu$-dependence 
it becomes unphysical for $\mu \ll m_Q$,
\beq 
\bar m_Q(\mu) \simeq \bar m_Q(\bar m_Q)\left[ 
1+ \frac{2\alpha_s}{\pi} {\rm log}\frac{m_Q}{\mu} \right] 
\label{msbarlog}
\eeq
where actual running becomes much slower. 
The $\overline {\rm MS}$ mass is thus inappropriate for treating 
decay processes where the relevant scales are well below $m_Q$ and the 
evolution to a low normalization point $\mu \ll m_Q$ becomes 
crucial.

The kinetic mass in perturbation theory order by order is defined 
by subtracting 
from the pole mass the perturbative contributions of 
the heavy quark parameters:
\beq
m_b(\mu)=m_b^{pole} - \left[\overline{\Lambda}(\mu)\right]_{\rm pert}
-\frac{\left[\mu_{\pi}^2(\mu)\right]_{\rm pert}}{2m_b(\mu)}
%%% - \frac{\left[\rho_D^3(\mu)\right]_{\rm pert}
%%% -\left[\rho^3(\mu)\right]_{\rm pert}}{4m_b(\mu)^2}- \ldots .
\label{mrunn}
\eeq
where the latter are determined from the corresponding SV sum rules 
cut at energy $\epsilon\!=\!\mu$:
\beq 
\bar\Lambda(\mu)_{\rm pert}= \frac{4}{3}C_F  
\frac{\alpha_s(M)}{\pi} \mu 
\left(1+\frac{\as}{\pi} \left[\frac{\beta_0}{2}
\left(\ln{\frac{M}{2\mu}} + \frac{8}{3}\right)-
C_A\left(\frac{\pi^2}{6}-\frac{13}{12}\right)\right]+...\right)\;,
\label{lamper}
\eeq
with the standard notation $\bar \Lambda \equiv \lim_{m_Q \!\to\! \infty} 
[M_{H_Q}\!-\!m_Q]$, \,and 
\beq
\mu_{\pi}^2 (\mu)_{\rm pert} = 
C_F 
\frac{\alpha_s(M)}{\pi}\mu^2
\left(1+\frac{\as}{\pi}\left[\frac{\beta_0}{2} 
\left(\ln{\frac{M}{2\mu}}+\frac{13}{6}\right)-
C_A\left(\frac{\pi^2}{3}-\frac{13}{12}\right)\right]+...\right)
 \; ,
\label{mupert} 
\eeq
where $C_F\!=\!\frac{4}{3}$ and $C_A\!=\!N_c\!=\!3$, and $M$ denoting an
arbitrary scale used to normalize $\alpha_s$. The latter
is assumed in the $\overline{{\rm MS}}$ scheme in Eqs.~(\ref{lamper}) and 
(\ref{mupert}). The thus defined mass exhibits only a linear in $\mu$
dependence on the scale,
\beq 
\frac{{\rm d}m_Q^{kin}(\mu)}{{\rm d}\mu} = 
-\frac{16}{9}\frac{\alpha_s(\mu)}{\pi} - 
\frac{4}{3} \frac{\alpha_s(\mu)}{\pi} \frac{\mu}{m_Q} 
+ {\cal O}\left( \alpha_s^2\right) 
\; , 
\eeq 
compared to Eq.~(\ref{msbarlog}). The pole mass would 
corresponds to integrating this evolution down to $\mu\!=\!0$ 
at each particular order -- alas it leads to an ill-defined result.

One could have extended the definition shown in Eq.~(\ref{mrunn}) to
include the corrections scaling like $\alpha_s^k \frac{\mu^3}{m_Q^2}$
and higher. Besides the known perturbative contribution in $\rho_D^3$ this
requires the similar terms for the nonlocal expectation values
$\rho_{\pi\pi}^3$ and $\rho_S^3$; the second-order non-BLM pieces
there have not been calculated. The one-loop effects, on the other
hand, appear to cancel in the sum of all $\mu^3/m_Q^2$ terms, and the
resulting shift beyond one loop (say, including evaluated BLM
corrections) is tiny. In any case for the beauty mass this makes no
visible difference whatsoever. There is an additional more subtle
reason not to pay much attention to such terms. Therefore, we prefer 
to consider Eq.~(\ref{mrunn}) as exact defining what is understood by 
the running masses, unless accuracy in their practical extraction will
improve in the future by orders of magnitude.

Often ``HQET parameters'' like 
$\La$, $-\lambda_1$ etc. are quoted rather than 
$\La(\mu)$, $\mu^2_{\pi}(\mu)$ with appropriate $\mu$; they
simply correspond to  taking  the limit $\mu \!\to\! 0$, and in this
respect are the close relatives of the pole masses.
Likewise they are not well-defined; moreover, the irreducible 
uncertainty and order-by-order instability in their values is of order
unity compared to their magnitude. In concrete calculations it is often
possible to identify their relation to well-defined Wilsonian
expectation values.
Yet one has to be aware that their
values depend sensitively on the  context in which they appear, and
one cannot transfer them from  one case to the next blindly. As a rule of
thumb one has for routinely used first-order calculations with the
second-order BLM improvement using $\alpha_s(m_b)$ coupling equated to
$0.22$
\beq 
\bar \Lambda_{\rm HQET}\simeq \bar \Lambda (1\; {\rm GeV}) - 
0.255\GeV\,, \qquad 
- \lambda _1 \simeq  \mu _{\pi}^2 (1\GeV) - 0.18 (\GeV)^2 \, .
\label{transl} 
\eeq

Finally we quote the relation between the running $m_Q(\mu)$ and
$\overline{\rm MS}$ mass $\bar{m}_Q(m_Q)$ including full $\alpha_s^2$
and third-order BLM corrections: 
{\small \bea
\nonumber
m_b(\mu)  &\msp{-5.1}=\msp{-5.1}& 
\overline{m}_b(\overline{m}_b)\left\{1+\frac{4}{3} 
\frac{\asMS(m_b)}{\pi}\left(1- \frac{4}{3} \frac{\mu}{\overline{m}_b} -
    \frac{\mu^2}{2\overline{m}_b^2} \right)\right. \\
\nonumber
&& \msp{-4}+\left(\frac{\asMS (m_b)}{\pi}\right)^2 \left[\frac{\beta_0}{2}
  \left(\frac{\pi^2}{6}+\frac{71}{48}\right) + \frac{665}{144} +
  \frac{\pi^2}{18}\left(2\ln{2} - \frac{19}{2}\right)
 - \frac{1}{6} \zeta (3) - \frac{8}{3}\right. \\
\nonumber
&& \msp{7.8} - 
\frac{\mu}{m_b}\left(
  \frac{8 \beta_0}{9} \left(\ln{\frac{m_b}{2\mu}}\!+\!
\frac{8}{3}\right) -
  \frac{8\pi^2}{9} + \frac{52}{9}\right)\left. \left. 
-\frac{\mu^2}{m_b^2}\left( \frac{\beta_0}{3} 
\left(\ln{\frac{m_b}{2\mu}}+ \frac{13}{6}\right) -
  \frac{\pi^2}{3} + \frac{23}{18}\right)\right]\right. 
\\ \nonumber 
&& \msp{-4}
+ \left(\frac{\beta_0}{2}\right)^2 \left(\frac{\alpha_s}{\pi}\right)^3
\left\{\frac{2353}{2592}+\frac{13}{36}\pi^2 +\frac{7}{6} \zeta(3)-
\frac{16}{9}\frac{\mu}{m_b}\left[\left(\ln{\frac{m_b}{2\mu}} \!+\! 
\frac {8}{3}\right)^2+
\frac{67}{36}-\frac{\pi^2}{6}\right]
\right. 
\\
&&\msp{30}
\left. 
- \;
\frac{2}{3} \frac{\mu^2}{m^2} 
\left[\left(\ln{\frac{m_b}{2\mu}} \!+\! \frac {13}{6}\right)^2+
\frac{10}{9}-\frac{\pi^2}{6}\right]
\right\} 
\;.
\label{mrunnMS}
\eea}
\hspace*{-.2em}This can serve as 
a useful reference to translate between the kinetic
and other possible low-scale running masses, along with the similar
relation between the pole and kinetic masses.

\subsection{Perturbative corrections in the Wilsonian approach}
\label{ap2}

Here we give relations allowing one to determine in practice perturbative
coefficients in the approach employing the Wilsonian separation of
large and small virtualities.

The starting point is the general short-distance expansion exemplified
by Eq.~(\ref{10}) for $\gsl$, or its practically employed 
form Eq.~(\ref{SLwid}). As stressed in Sect.~\ref{GAMMA}, it must be
$\mu$-independent. Then one considers it in perturbation theory
limited in practice to a particular order in $\alpha_s$ and
$1/m_b$. Requiring $\mu$-independence completely defines the
$\mu$-dependence of the Wilson coefficients once the $\mu$-dependence of
masses and matrix elements in perturbation theory is known 
\cite{optical,blmope}. For
truncated perturbation theory the actual expression has a residual
dependence on $\mu$ to the uncalculated orders in $\alpha_s$. On the
other hand, in the finite-order perturbation theory one can formally
put $\mu\!=\!0$. This would set all the power-suppressed expectation values
to zero and yield pole masses, hence fixing the perturbative series
for $A^{\rm pert}(r;0)$ to be the `conventional' perturbative
corrections in the pole mass scheme. This provides the initial
condition for perturbative $\mu$-dependence at $\mu\!=\!0$.

To give the practical implementation for $\gsl$ we need to specify
more precisely our definition in Eq.~(\ref{80}): now we write
\beq
z_0(r)\, A^{\rm pert}(r;\mu)=z_0(r)+a_1(r;\mu)\mbox{$\frac{\alpha_s}{\pi}$}+ 
a_2(r;\mu)\left(\mbox{$\frac{\alpha_s}{\pi}$}\right)^2 +
a_3(r;\mu)\left(\mbox{$\frac{\alpha_s}{\pi}$}\right)^3 + ...
\label{380}
\eeq
and use the short-hand notations $a_k^{(0)}(r)\equiv a_k(r,0)$
for the perturbative coefficients at $\mu\!=\!0$ (pole mass
scheme). Then one has %%% \note{add effect of Darwin term}
\bea
a_1(r;\mu)\msp{-3} &=& \msp{-3} 
a_1^{(0)}(r) + \left[5\Lambda_1+3p_1 \right]\,z_0(r)- 
\rho_1\,D(r)+
\left[2(\sqrt{r}\!-\!r)\Lambda_1+(1\!-\!r)p_1 \right]\,
\frac{{\rm d}z_0(r)}{{\rm d}r}\label{a1}
\\
\nonumber
a_2(r;\mu)\msp{-3} &=& \msp{-3} 
a_2^{(0)}(r) + \left[5\Lambda_1+3p_1 -\rho_1
\mbox{$\frac{D(r)}{z_0(r)}$}\right]\,
a_1^{(0)}(r)
+\left[2(\sqrt{r}\!-\!r)\Lambda_1+(1\!-\!r)p_1 \right]\,
\frac{{\rm d}a_1^{(0)}(r)}{{\rm d}r}
\\
\nonumber
&& \msp{-12}
+ \left[5\Lambda_2+3p_2 +10\Lambda_1^2+4p_1^2 +
\frac{25}{2}\Lambda_1 p_1-\left(\rho_2+ 
\rho_1 (5\Lambda_1+\mbox{$\frac{7}{2}$} p_1)\right)
\mbox{$\frac{D(r)}{z_0(r)}$}\right]\,z_0(r) +
\mbox{$\frac{\rho_1^2 D^2(r)}{z_0(r)}$}\\
\nonumber
&& \msp{-12}
+\left[2(\sqrt{r}\!-\!r)\Lambda_2+(1\!-\!r)p_2+
(1\!+\!6\sqrt{r}\!-\!7r)\Lambda_1^2 
+(\frac{1}{4r}\!+\!2\!-\!\frac{9}{4}r)p_1^2
\:+ \right. \\ \nonumber
&& \msp{-6} %% \left.
(\frac{1}{\sqrt{r}}\!+\left.\!3\!+\!4\sqrt{r}\!-\!8r)\Lambda_1 p_1
\rule{0mm}{5mm} \right]\,\frac{{\rm d}z_0(r)}{{\rm d}r} 
+ \frac{1}{2}
\left[2(\sqrt{r}\!-\!r)\Lambda_1+(1\!-\!r)p_1 \right]^2
\frac{{\rm d}^2z_0(r)}{{\rm d}r^2}\\
&& \msp{-6} -\rho_1\left[2(\sqrt{r}\!-\!r)\Lambda_1+(1\!-\!r)p_1 \right]
\,\frac{{\rm d}D(r)}{{\rm d}r}
\,,\qquad
\label{a2}
\eea
and the coefficient $a_1^{(0)}(r)$ is given by \cite{muon,analyt} 
\bea
\nonumber
a_1^{(0)}(r)&\msp{-5}=\msp{-5}&
\frac{1}{24}\left[75 - 12\,{\pi }^2 - 956\,r -
192\,{\pi }^2\,r^2 + 956\,r^3 - 3\,\left( 25 + 4\,{\pi }^2 \right)
\,r^4 \right. \\
\nonumber
&&\msp{-9} +384\,{\pi }^2\,r^{\frac{3}{2}}\,\left( 1 + r \right)+
4\,\left( -17 + 64\,r - 64\,r^3 + 17\,r^4 \right) \,\ln (1 - r)\\
\nonumber
&& \msp{-9}-4\,r\,\left( 60 + 270\,r - 4\,r^2 + 17\,r^3 -
384\,{\sqrt{r}}\,\left( 1 + r \right) \,\ln (1 + {\sqrt{r}}) \right)
\, \ln (r) \\
\nonumber
&&\msp{-9} +48\,\left( 1 \!-\! 16\,r^{\frac{3}{2}} + 30\,r^2 \!-\! 
16\,r^{\frac{5}{2}}
+ r^4 \right) \,\ln{(1 \!-\! r)}\,\ln{r} -  
12\,r^2\,\left( 36 + r^2 \right) \,\ln^2{r}\\
\nonumber
&&\msp{-9} - \left. 3072\,r^{\frac{3}{2}}\,\left( 1 + r \right) 
\,{\rm Li}_2(\sqrt{r}\,) + 
18\,\left( 4 + 64\,r^2 + 4\,r^4 + \frac{128\,r^{\frac{3}{2}}\,\left( 1
+ r \right) }{3} \right) \,{\rm Li}_2(r)\right]. 
\eea
The combinations $\Lambda_{1,2}$, $p_{1,2}$ and $\rho_{1,2}$ are the 
coefficients of the perturbative expansion of 
$\frac{1}{m_b}\left[\Lambda(\mu)\right]_{\rm pert}$
and $\frac{1}{m_b^2}\left[\mu_{\pi}^2(\mu)\right]_{\rm pert}$ 
in Eqs.~(\ref{lamper}) and (\ref{mupert}) and 
of  $\frac{1}{m_b^3}\left[\rho_{D}^3(\mu)\right]_{\rm pert}$, 
to first and second orders in
$\alpha_s$, respectively: 
\bea
\label{311}
\Lambda_1\msp{-3} &=& \msp{-3}\frac{4}{3}C_F\frac{\mu}{m_b}, \qquad
\Lambda_2=\Lambda_1\left[\frac{\beta_0}{2}\left(\ln{\frac{M}{2\mu}}+\,
\frac{8}{3}\,\right) 
- C_A\left(\frac{\pi^2}{6}\!-\!\frac{13}{12}\right)\right]\,, \\
p_1\msp{-3} &=& \msp{-3} \;\;C_F\frac{\mu^2}{m_b^2}, \qquad \;
p_2=\,p_1 \left[\frac{\beta_0}{2}\left(\ln{\frac{M}{2\mu}}+
\frac{13}{6}\right) 
- C_A\left(\frac{\pi^2}{6}\!-\!\frac{13}{12}\right)\right]\,, \\
\rho_1\msp{-3} &=& \msp{-3} \frac{2}{3}C_F\frac{\mu^3}{m_b^3}, \qquad \;
\rho_2=\,\rho_1 \left[\frac{\beta_0}{2}\left(\ln{\frac{M}{2\mu}}+
\,2\,\right) 
- C_A\left(\frac{\pi^2}{6}\!-\!\frac{13}{12}\right)\right]. \qquad \qquad
\label{312}
\eea

It should be noted that we did not include here shifts due to 
the spin-singlet 
$1/m_b$ pieces of the kinetic and chromomagnetic operators (we have
computed them to order $\alpha_s$ and to all orders in BLM). These
corrections are $1/m_b^3$ and governed by the full $1/2m_b$ scale,
hence totally insignificant in practice. We did include the effect
of the Darwin operator to two full loops (modulo uncalculated ${\cal
O}(\alpha_s)$ piece $A_D^{\rm pert}$), since its
coefficient is enhanced and strongly dominates over the $1/m_b^3$ effects.

The improvement by the Wilsonian cutoff of the resulting
perturbative series can be illustrated by comparing them at
$\mu\!=\!1\GeV$
\beq 
A_{\rm pert}(1\GeV)\simeq
1 - 0.94 \frac{\alpha_s(m_b)}{\pi} - 
5.1 \left( \frac{\alpha_s(m_b)}{\pi}\right)^2 - 
17 \left( \frac{\alpha_s(m_b)}{\pi}\right)^3 + 
63 \left( \frac{\alpha_s(m_b)}{\pi}\right)^4 +...
\label{pertser}
\eeq
with those in the `pole' (HQET) scheme:
\beq 
A_{\rm pert}^{\rm pole}\simeq
1 - 1.78 \frac{\alpha_s(m_b)}{\pi} - 
15.8 \left( \frac{\alpha_s(m_b)}{\pi}\right)^2 - 
230 \left( \frac{\alpha_s(m_b)}{\pi}\right)^3 - 
3640 \left( \frac{\alpha_s(m_b)}{\pi}\right)^4 -...
\label{pertserpole}
\eeq 
where numbers refer to $m_c/m_b=0.25$.

\subsection{BLM summation with Wilsonian cutoff}
\label{ap3}

The BLM correction in practice are simpler to calculate directly,
using the generalized order-$\alpha_s$ relation (\ref{a1}) between Wilson
coefficients with and without the cutoff. The technique of BLM
resummation was recently reviewed in this context in
Ref.~\cite{blmvcb}. It requires computing one-loop corrections with a
non-zero gluon mass $\lambda$. 
The expressions for the one-loop correction to the perturbative width
with non-zero gluon mass are rather lengthy and can be found in
Ref.~\cite{bbbsl}. One also needs similar terms in 
$\left[\La(\mu)\right]^{\rm pert}$,
$\left[\mu_{\pi}^2 \right]^{\rm pert}$ and 
$\left[\rho_D^3 \right]^{\rm pert}$.
These are calculated by integrating the SV spectral
density of Ref.~\cite{blmope} with the proper power of energy and
are given as follows: 
\bea
\nonumber
\Lambda_1(\mu;\,\lambda)\msp{-3}&=&\msp{-3} 
\frac{16\alpha_s}{9\pi}\,\frac{1}{m_b}\;
\vartheta \,(\mu^2\!-\!\lambda^2)\,\left[
(1\!-\!\mbox{$\frac{\lambda^2}{4\mu^2}$})\sqrt{\mu^2\!-\!\lambda^2}
 \!-\!\frac{3\pi}{8}\lambda+ 
\frac{3}{4}\lambda 
\arcsin\frac{\lambda}{\mu}\right]\,,\\ \nonumber
p_1(\mu;\,\lambda)\msp{-3}&=&\msp{-3} 
\;\frac{4\alpha_s}{3\pi}\,\frac{1}{m_b^2}\;\,
\vartheta \,(\mu^2\!-\!\lambda^2) \;
\frac{(\mu^2\!-\!\lambda^2)^{3/2}}{\mu}\,,\\
\rho_1(\mu;\,\lambda)\msp{-3}&=&\msp{-3} 
\frac{8\alpha_s}{9\pi}\,\frac{1}{m_b^3}\;
\vartheta \,(\mu^2\!-\!\lambda^2)\,\left[\sqrt{\mu^2\!-\!\lambda^2}
(1\!+\!\mbox{$\frac{\lambda^2}{2\mu^2}$})\!-\!\frac{3\lambda^3}{2}\left(
\frac{\pi}{2}-\arcsin{\frac{\lambda}{\mu}}
\right)
\right]\,.
\label{HQPlambda}
\eea 
One then has
\bea
a_1(r;\mu;\lambda) &\msp{-3} = \msp{-3}& 
a_1^{(0)}(r;\lambda) +
\left[5\Lambda_1(r;\mu;\lambda)+3p_1(r;\mu;\lambda) 
\right]\,z_0(r)- 
\rho_1(r;\mu;\lambda)\,D(r) \nonumber \\
& & \msp{15}
+ \left[2(\sqrt{r}\!-\!r)\Lambda_1(r;\mu;\lambda)+(1\!-\!r)p_1(r;\mu;\lambda)
\right]\,
\frac{{\rm d}z_0(r)}{{\rm d}r}
\label{a1mulam}
\eea

The perturbative coefficients are explicitly given by the series
\bea
\nonumber
A^{\rm BLM}(r;\mu)\msp{-3}& =& \msp{-3}1\;+\; 
a_1(r;\mu;0)\frac{\alpha_s(m_b)}{\pi}
\;+\;
\sum_{n=0}^{\infty}\: 
\mbox{$\frac{4}{\beta_0}
\left(\frac{\beta_0\alpha_s(m_b)}{4\pi}\right)^{n+2}$} \times
\;\\
& & \msp{-27}
\sum_{k=0}^{\frac{n}{2}}\, 
(-\pi^2)^k \;\mbox{\large$\raisebox{-.5mm}{$C$}_{_{\,n+1}}^{^{\,2k+1}}$}
\cdot\int
\frac{{\rm d}\lambda^2}{\lambda^2}\,\left[\ln{\frac{m_b^2}{\lambda^2}}
\!+\!\frac{5}{3}\right]^{n-2k}
\!\!\mbox{$\left(a_1(r;\mu;0)\mbox{\large$\frac{m_b^2}{m_b^2+ 
e^{\mbox{\tiny -5/3}}\lambda^2}$}\!-\! 
a_1(r;\mu;\lambda^2) \right)$},
\label{424}
\eea
and the resummed result reads as
\bea
\nonumber
A^{\rm BLM}(r;\mu)\msp{-3}& =& \msp{-3} 1+
a_1(r;\mu;0)\frac{\alpha_s(m_b)}{\pi}\\
\nonumber
& & \msp{-17}+
\int_{-\infty}^{\infty} \; {\rm d}t\:  
\frac{\frac{\beta_0}{4}\,\left(\frac{\alpha_s}{\pi}\right)^2}{\left(1+ 
\frac{\beta_0\alpha_s}{4\pi}(t\!-\!\frac{5}{3})\right)^2+
\left(\frac{\beta_0}{4}\alpha_s\right)^2} 
\:\left( a_1(r;\mu;0)\frac{1}{1+e^{t-\frac{5}{3}}}\!-\! 
a_1(r;\mu;e^{t}m_b^2)\right)
\\
& & \msp{2}
- \frac{4}{\beta_0}\left[\frac{m_b^2}{m_b^2\!-\!\Lambda_V^2}a_1(r;\mu;0)
-a_1(r;\mu;-\Lambda_V^2)
\right]
\;,
\label{420}
\eea
with
\beq
\Lambda_V^2\,=\,m_b^2 \:e^{-\frac{4\pi}{\beta_0\alpha_s(m_b)}+\frac{5}{3}}\;.
\label{422}
\eeq
Numerically at $\,m_c/m_b\!=\!0.25$, $\;\mu/m_b\!=\!1/4.6\;$ and 
$\;\alpha_s(m_b)\!=\!0.22\;$ we obtain  
\beq
A^{\rm BLM}(r;\mu)\,=\, 0.915\;.
\label{426}
\eeq

\end{document}